\documentclass[preprint, 3p, twocolumn]{elsarticle}
\usepackage{tabularx}
\usepackage{graphicx}
\usepackage{float}
\usepackage{color}
\usepackage{rotating}
\usepackage{amssymb}
\usepackage{siunitx}
\usepackage{xcolor,colortbl}
\usepackage{todonotes}
\usepackage[numbers]{natbib} 
\usepackage[utf8]{inputenc}
\usepackage{doi} 
\usepackage{adjustbox}	
\usepackage{mathtools}
\usepackage{wasysym}
\definecolor{Gray}{gray}{0.85}

\newcolumntype{a}{>{\columncolor{Gray}}c}
\newcolumntype{b}{>{\columncolor{white}}c}

\begin{document}
\begin{abstract}

Neutron scattering is a well-established tool for the investigation of the static and dynamic properties of condensed matter systems over a wide range of spatial and temporal scales. Many studies of high interest, however, can only be performed on small samples and typically require elaborate environments for variation of parameters such as temperature, magnetic field and pressure. To improve the achievable signal-to-background ratio, focusing devices based on elliptic or parabolic neutron guides or Montel mirrors have been implemented. Here we report an experimental demonstration of a nested mirror optics (NMO), which overcomes some of the disadvantages of such devices. While even simpler than the original Wolter design, our compact assembly of elliptic mirrors images neutrons from a source to a target, minimizing geometric aberrations, gravitational effects and waviness-induced blurring. Experiments performed at MIRA at FRM-II demonstrate the expected focusing properties and a beam transport efficiency of $\SI{72}{\percent}$ for our first prototype. NMO seem particularly well-suited to i) extraction of neutrons from compact high-brilliance neutron moderators, ii) general neutron transport, and iii) focusing and polarizing neutrons. The phase space of the neutrons hitting a sample can be tailored on-line to the needed experimental resolution, resulting in small scattering backgrounds. As additional benefits, NMO situated far away from both the moderator and the sample are less susceptible to radiation damage and can easily be replaced. NMO enable a modular and physically transparent realization of beam lines for neutron physics similar to setups used in visible light optics.
\end{abstract}

\begin{keyword} focusing devices, neutron transport, neutron scattering, polarized neutrons, supermirror
\end{keyword}

\begin{frontmatter}

\title{Nested mirror optics for neutron extraction, transport, and focusing}

\author[TUM]{Christoph Herb\corref{Cor}}
\ead{christoph.herb@frm2.tum.de}
\cortext[Cor]{Corresponding author. +49 89 289 14315}
\author[ILL]{Oliver Zimmer}
\author[TUM,FRM]{Robert Georgii}
\author[TUM]{Peter B\"oni}

\address[TUM]{Physics Department E21, Technical University of Munich, D-85748 Garching, Germany}
\address[ILL]{Institut Laue-Langevin, 71 avenue des Martyrs, F-38042 Grenoble, France}
\address[FRM]{Heinz Maier-Leibnitz Zentrum, Technical University of Munich, D-85748 Garching, Germany}

\date{\today}

\end{frontmatter}



\section{Introduction}
\label{sect:introduction}

Many researchers active in a vast range of scientific disciplines appreciate neutron scattering as an important tool to investigate static and dynamic properties of condensed matter systems across large spatial and temporal scales. While able to address a plethora of scientific topics, experiments often suffer from the limited neutron flux available from present-day sources. The most interesting experiments can often be performed only on small samples, such as, for instance, laboratory grown single crystals of novel, complex materials or of biological macromolecules \cite{shiraneNeutronScatteringTripleAxis2002}. Further limitations on sample size are imposed by the experimental conditions needed to address a particular research problem. For example, continued progress in the investigation of phase transitions, which are key to understanding the physics of quantum criticality and superconductivity \cite{guoPressureDrivenQuantumCriticality2012, kobayashiPressureinducedSuperconductivityFerromagnet2002, pfleidererSuperconductingPhasesElectron2009}, is made possible by subjecting samples to extreme temperatures, magnetic fields, and pressures. Most notably, ultra-high pressures are achievable only with tiny samples. Advanced sample environments furthermore decrease neutron access to the sample, e.g., requiring neutrons to pass through structural materials needed to maintain sample temperatures, further limiting signal-to-background.\\

To some extent, these difficulties can be mitigated by focusing of the incident neutron beam, which can increase the scattering signal from small samples. This may also improve the background, provided that one succeeds to prepare a clean beam of only those neutrons that may contribute to the scattering signal. Avoiding the ``halo'' of useless neutrons originating from, e.g., scattering in intensely exposed apertures near the sample region is also beneficial for an improved signal-to-background ratio.\\

It is the main goal of this paper to bring to the attention of neutron instrument developers a first experimental demonstration of a new type of neutron transport device, with properties that are appealing for a variety of applications. The paper is organized as follows: after this Introduction, Section \ref{sect:operationalPrinciple} presents the operational principle of elliptical nested mirror optics (NMO), along with basic implementations; Section \ref{sect:experimentalSetup} describes a prototype device and an experimental setup for its characterization with neutrons; Section \ref{sec:meas_res} presents experimental results, which are compared to simulations in Section \ref{sect:comp_exp_sim}; Section \ref{sect:applications} discusses various applications of NMO. The paper concludes with a summary and an outlook, and is complemented by two Appendices.\\


\section{Operational principle and basic implementations of nested elliptical mirror optics}
\label{sect:operationalPrinciple}

Due to their lack of an electric charge, guiding and focusing of neutrons by mirror reflection from surfaces is presently the predominant method to prepare neutron beams for research. After the discovery of the total reflection of neutrons at small angles from metal-coated surfaces in 1945 by Fermi and Zinn \cite{fermiReflectionNeutronsMirrors1946}, Maier-Leibnitz and Springer invented neutron guides with a rectangular cross section, to overcome the decrease of the neutron flux by the squared inverse of the distance from a point-like source \cite{maier-leibnitzUseNeutronOptical1963}.\\

Early neutron guides frequently used nickel coatings, which, compared with other materials, possess a particularly large ``critical angle'', below which neutrons of a given wavelength, $\lambda$, undergo total reflection from the surface. For this reason, this material-dependent quantity is frequently discussed using nickel as a reference, and is written
\begin{equation}\label{eq:crit_angle}
\theta_{\text{c}, m} = m \kappa \lambda,
\end{equation}
where $\kappa = \SI{0.099}{deg/\angstrom} = {0.00173}\,\si{rad/\angstrom}$ with a numerical value chosen to be characteristic for Ni with natural isotopic composition. The dimensionless factor $m$, referred to as the ``$m$-value'' of a mirror coating, quantifies the deviation of the critical angle from the critical angle of Ni, i.e., $\theta_{\text{c}, m} = m \theta_{\text{c}, 1} = m\theta_{\text{c}}(\text{Ni})$. Later on, the development of supermirrors \cite{mezeiNovelPolarizedNeutron1976a, mezeiPolarizingSupermirrorDevices1992} has enabled a significant increase of the critical angle by up to $m \simeq 8$ times \cite{schanzerNeutronOpticsApplications2016}. As a result, compared to Ni-coated surfaces, supermirror coated guides can transport neutrons emanating from ambient and even from hot moderators within a significantly increased solid angle, or ``beam divergence'', defined as twice the critical angle per lateral dimension.\\

\begin{figure}[htb]
\centering
\includegraphics[width=\linewidth]{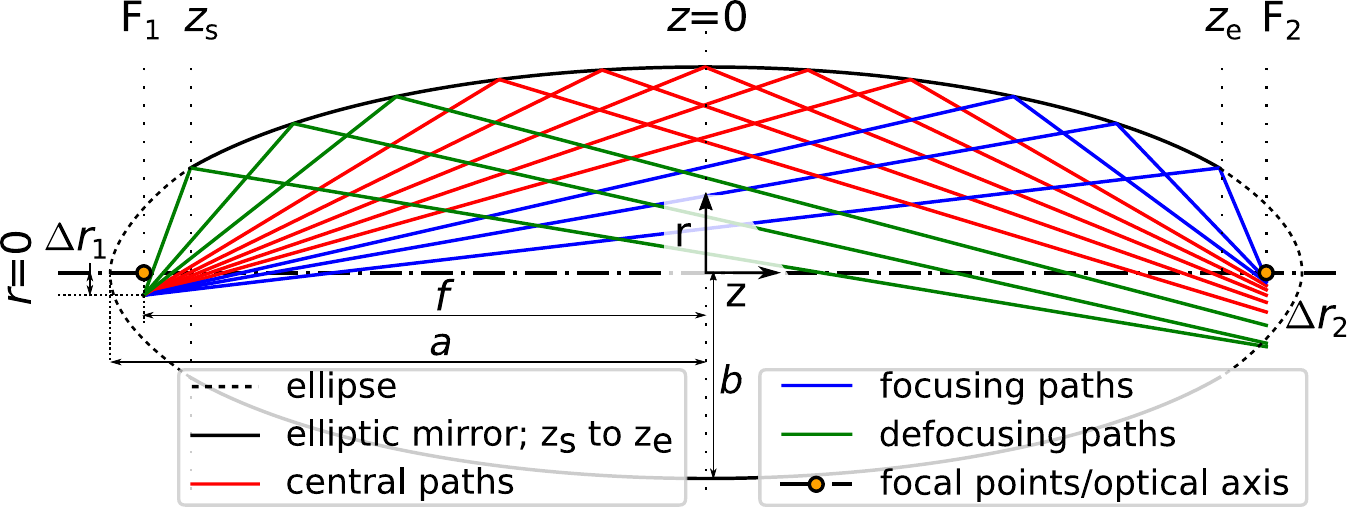}
\caption{Flight paths of neutrons emitted near the first focal point F$_1$ of an elliptic mirror with a negative offset $\Delta r_1$ from the optical axis (shown as broken black line). The semi-major and semi-minor axes of the ellipse, $a$ and $b$, respectively, define the focal points, located at $z = \pm f$, where $f=\sqrt{a^2-b^2}$. Neutrons reflected at points with $z < 0$ arrive at F$_2$ with an increased offset, i.e., ``defocused'' ($\left|\Delta r_2\right| > \left|\Delta r_1\right|$, neutron paths indicated as green lines), whereas those reflected at $z > 0$ are focused toward F$_2$ ($\left|\Delta r_2\right| < \left|\Delta r_1\right|$, blue lines). Neutrons reflected at $z\approx 0$ arrive at F$_2$ with $\Delta r_2 \approx \Delta r_1$ (red lines).}
\label{fig:aberrations}
\end{figure}

In long neutron guides, losses due to imperfect mirror reflectivity can be reduced by using side walls with elliptical or parabolic shape \cite{aschauerNeutronGuidesFRMII2000, schanzerAdvancedGeometriesBallistic2004}, extending an earlier concept of a ``ballistic neutron guide'' with a diverging and reconverging cross section \cite{Mezei1997, Haese2002}. Consequently, most beam lines at the European Spallation Source (ESS) use elliptic guides to transport neutrons from the moderators to the instruments \cite{zendlerGenericGuideConcepts2015}. However, while such guides are well-suited to narrow beams, geometric aberrations strongly increase with source size \cite{rodriguezPropertiesEllipticalGuides2011}. As may be seen in the illustration of Fig.\,\ref{fig:aberrations}, unless reflection occurs within a small region near the co-vertex of the ellipse at $z=0$, defined as the origin of the optical axis, neutrons starting with vertical offset $\Delta r_1$ from the first focal point, F$_1$, will be either focused or ``defocused'' at the second focal point, F$_2$. The offset at F$_2$, $\Delta r_2$, is related to $\Delta r_1$ by \cite{rodriguezPropertiesEllipticalGuides2011}
\begin{equation}\label{eq:focdefoc}
\frac{\Delta r_2}{\Delta r_1} \approx \frac{f-z}{f+z}.
\end{equation} 
Hence, the aberrations at the sample position depend on the $z$-coordinate of the point of reflection, so that for a long elliptic guide strong variations of the divergence and intensity profile may occur at F$_2$, compared to the beam profile at F$_1$ \cite{Stahn2011}. Moreover, multiple reflections occur in the guide that reduce the overall transport efficiency \cite{cussenMultipleReflectionsElliptic2013}. In addition, a complete description of neutron focusing needs to consider the influence of gravity, which can also lead to a vertically distorted reflection pattern \cite{Weichselbaumer2015}.\\ \noindent

Geometric aberrations characterized by Eq.\,\ref{eq:focdefoc} can be limited by truncating the ellipse to a smaller region along the $z$-axis (compare Fig.\,\ref{fig:aberrations}). This modification, which keeps only the reflecting surfaces within a small region, for instance centered on $z=0$, would provide an unmagnified optical image of the primary beam, of width $\Delta r_2 \simeq \Delta r_1$. The shorter one chooses the length $l = z_\mathrm{e}-z_\mathrm{s}$ of this section, the better the definition of the beam image at F$_2$. However, short $l$ also results in small angular acceptance. 

\begin{figure}[htb]
\centering
\includegraphics[width = \linewidth]{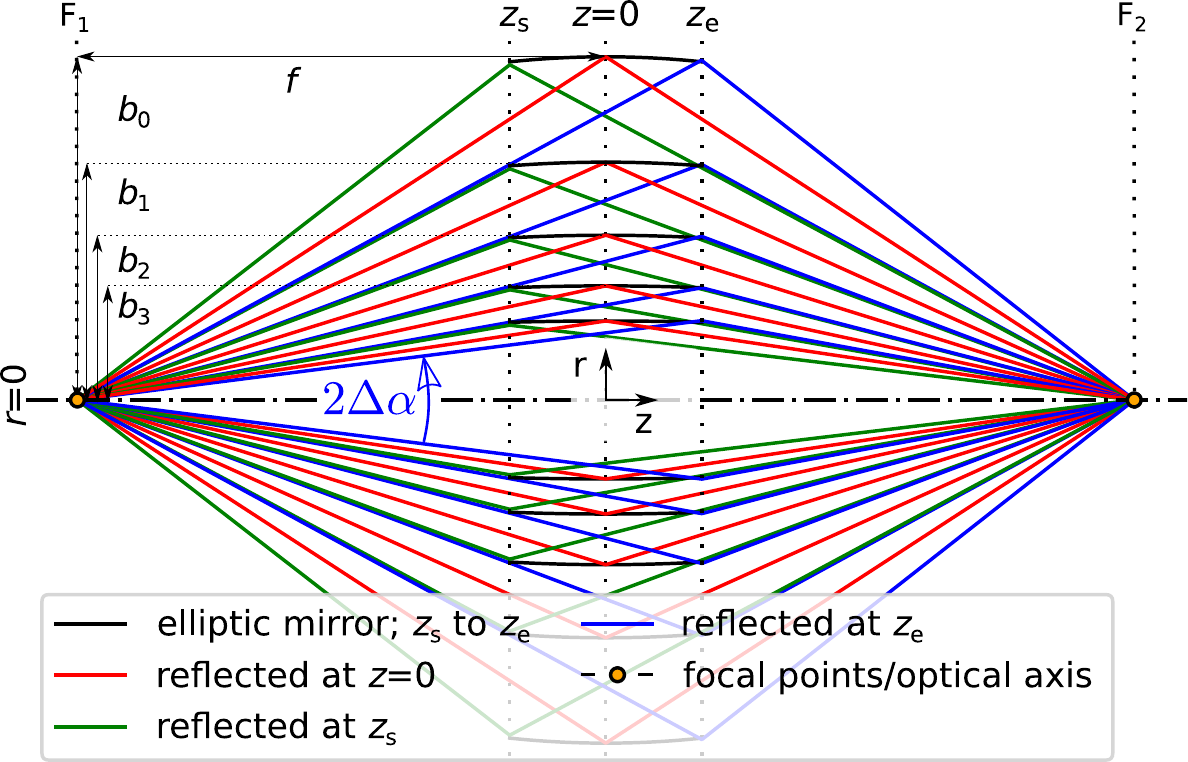}
\caption{Elliptic NMO including exemplary neutron beam paths. The black solid lines indicate the realized elliptic mirrors. Green and blue lines show the extreme neutron flight paths, being reflected at the entrance ($z_s$) and exit ($z_e$) of the NMO, respectively. The common focal points F$_1$ and F$_2$ of the elliptic mirrors are indicated by orange circles.}
\label{fig:nestedsketch}
\end{figure}

As discussed in \cite{zimmerImagingNestedmirrorAssemblies2018}, and shown in Fig.\,\ref{fig:nestedsketch}, by nesting several elliptic mirrors with common focal points but different semi-minor axes $b_n$, one can increase the divergence of the transported beam, and hence the total flux, without compromising the spatial beam definition. The maximum divergence covered by a nested mirror optics (NMO) is limited by the achievable $m$-value of the outermost supermirror. The semi-minor axis $b_0$ of the latter can be determined after fixing the locations of the two focal points. By requiring that a line originating from F$_1$ connects the back end of a specific mirror at $z = z_{\text{e}}$ with the front end of the adjacent inner mirror at $z = z_{\text{s}}$, one can construct a device capturing a geometrically defined range of angles in an iterative manner \cite{zimmerMultimirrorImagingOptics2016}. For a finite number of mirrors, rays with angles within a range $2 \Delta \alpha$, see Fig.\,\ref{fig:nestedsketch}, do not participate in the imaging process. The impact of this ``divergence hole'' on the neutron transport efficiency will be further studied in Section \ref{sec:ellipticNMO} where we analyze elliptic NMO in the context of applications. While one could fill the divergence hole with additional mirrors, an absorber centered on $z=0$ can remove the direct view onto the source along the optical axis, e.g., for suppression of fast-neutron background from a spallation source.\\

\begin{figure}[htb]
\centering
\includegraphics[width=\linewidth]{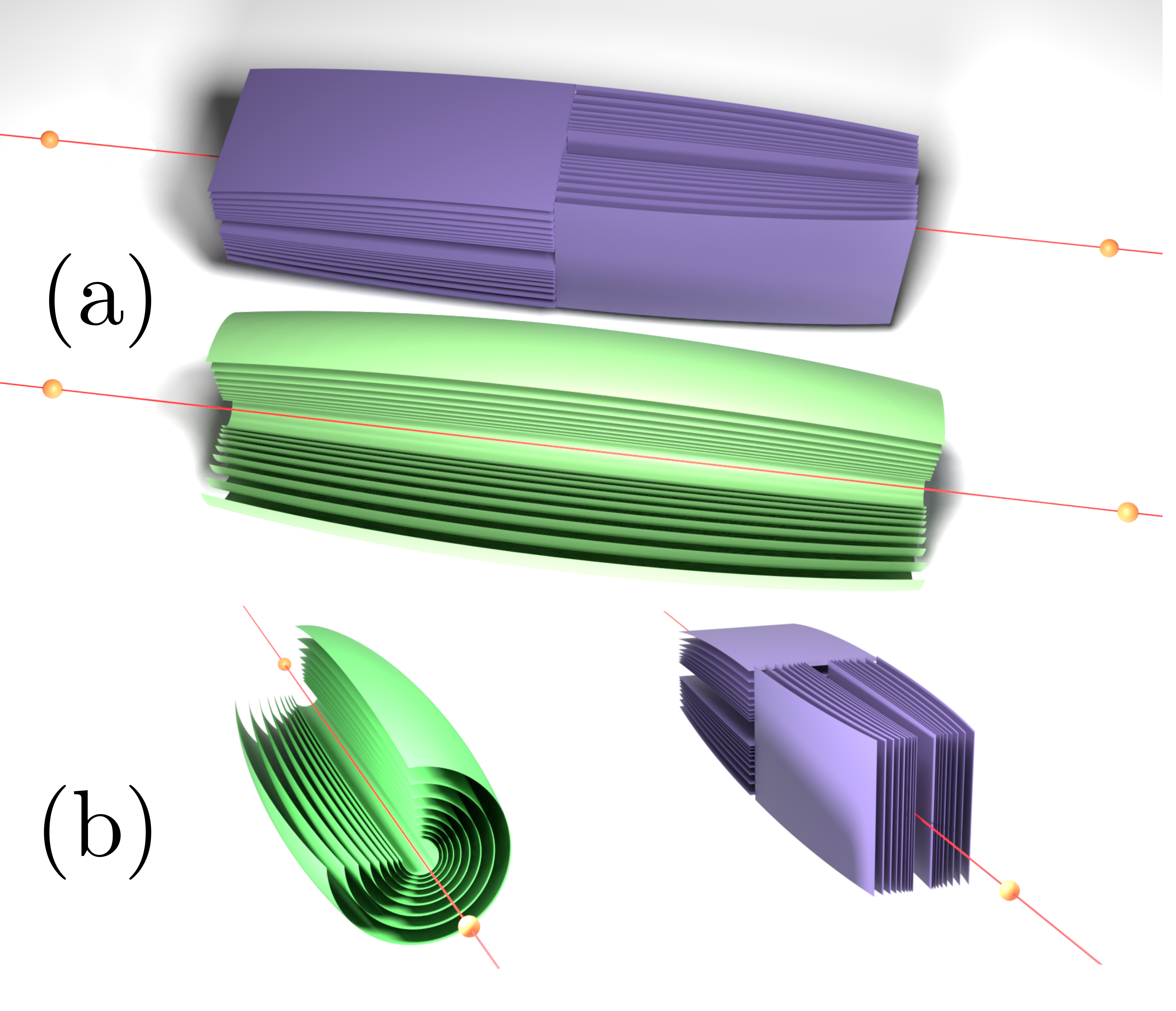}
\caption{Double-planar (violet) and toroidal (green) elliptic NMO. (a) and (b) show views from different angles to the optical axis. Representative focal points (orange spheres) are located on the optical axes shown in red. For the purpose of illustration, the curvature of the ellipses is vastly exaggerated and a quarter of the toroidal NMO is removed.}
\label{fig:doublefoc}
\end{figure}

In Fig.\,\ref{fig:nestedsketch}, $r$ can designate either a Cartesian coordinate, or the radial coordinate in a cylindrical geometry. In the latter case, the sketch represents a toroidal mirror system \cite{khaykovichXrayTelescopesNeutron2011}, generated by rotating the shown nested truncated elliptic lines around the optical axis. The sketched system would thus comprise five rotationally symmetric mirrors. In the case with a Cartesian $r$-coordinate, the sketch shows a section view of an NMO, which refocuses neutrons only in one of the two dimensions transverse to the beam. To achieve a complete imaging of a narrow divergent beam, similar to the toroidal system, two orthogonal planar NMO can be combined. Figure \ref{fig:doublefoc} shows a graphic representation of both, the toroidal version and of a double-planar NMO with two spatially separated subsystems located in the $z$-sections $(-l,\, 0)$ and $(0,\, l)$. A more compact, albeit technically more demanding, possibility would be to locate both subsystems within a common $z$-section $(-l/2,\, l/2)$, thus intersecting each other.\\

A double-planar system with horizontal and vertical mirrors shares its symmetry with the rectangular neutron guides currently used for neutron transport and is thus well-suited to be combined with them. On the other hand, toroidal mirror systems might ultimately be more efficient because they refocus neutrons in both dimensions simultaneously by only a single reflection. However, the deposition of high-quality supermirrors directly on curved substrates seems technologically difficult \cite{Wu2019}. In the limit of very short mirrors, where the elliptic curve can be approximated by a straight line, coating of flat substrates, subsequently bent into a final cylindrical geometry, might be a viable alternative if one can keep mirror waviness well under control. As a further alternative, one could also envisage a tessellation made of small pieces of flat mirrors as a design principle to construct NMO.\\

\section{Experimental setup}
\label{sect:experimentalSetup}

To investigate the overall performance of an NMO, we constructed a prototype of a planar, elliptic system with unit magnification and performed several experiments using the multi-purpose three-axis spectrometer MIRA at the Maier-Leibnitz-Zentrum (MLZ) \cite{georgiiMultipurposeThreeaxisSpectrometer2018}. The quality of the neutron optical image and the efficiency of neutron transport, i.e., how many neutrons entering the optics are recovered near its second focal point, were of primary interest. However, the NMO prototype was constructed from polarizing supermirrors, left over from a polarizer assembly, and we therefore also took the opportunity to investigate the polarizing properties of the prototype.\\

\begin{figure}[htb]
\centering
\includegraphics[width = \columnwidth]{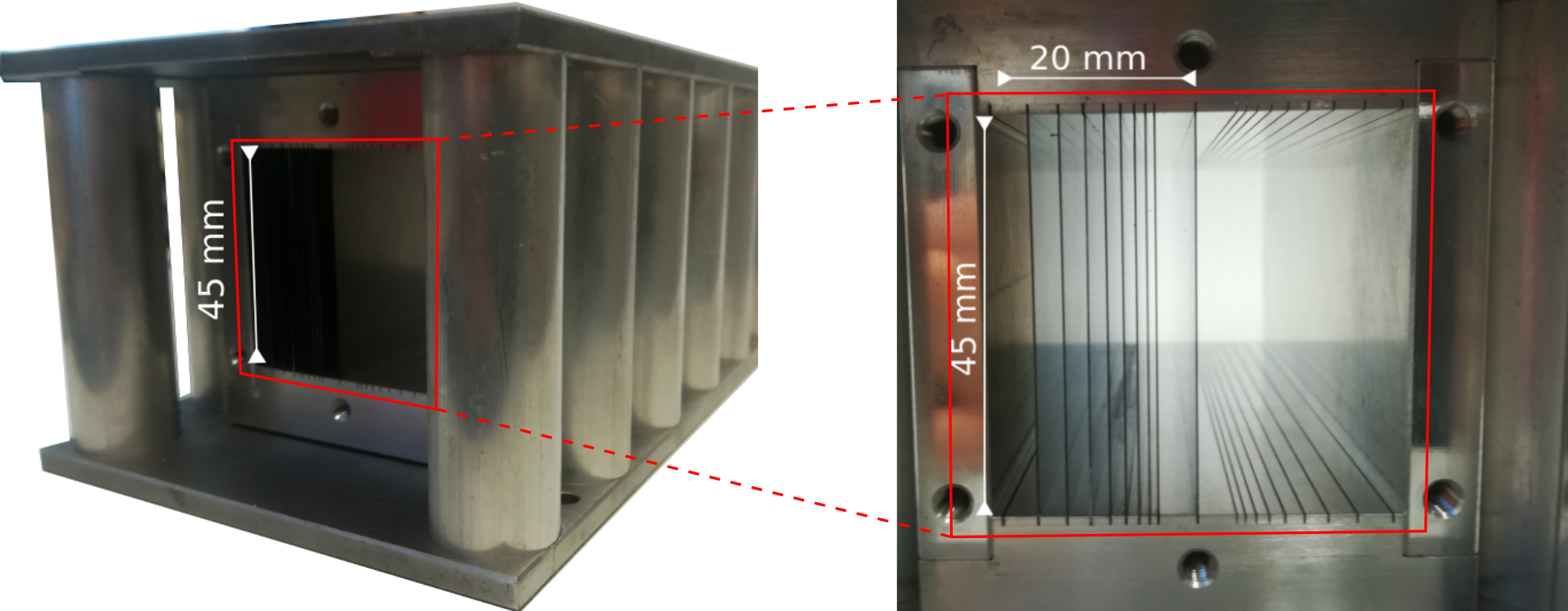}
\caption{Pictures of the NMO prototype with its eight elliptic mirror plates. A ninth, central and straight mirror solely facilitates the alignment of the optic. The grooves visible on the right hand side were left empty. The rods on the sides of the device contain the permanent magnets for magnetization of the supermirrors. (See text for a description of the geometry and more details of the mirror setup.)}
\label{fig:real_optic}
\end{figure}

Mono-crystalline silicon, coated on both sides with polarizing $m = 4.1$ FeSi supermirrors, were used as the mirror plates. As no absorbing sub-layer was used, the mirrors had a high transparency for neutrons exceeding the critical angle, $\theta_{\text{c}, m}$. Each plate had a substrate thickness of $d_{\text{sub}} = \SI{0.15}{mm}$, a length of $l = \SI{120}{mm}$, and a height of $\SI{45}{mm}$. Using the formulas presented in Ref.\,\cite{zimmerMultimirrorImagingOptics2016}, we designed an NMO consisting of eight such plates, with each of the two focal points situated at distance $f = \SI{600}{mm}$ from the center of the assembly. Note that, in the design process, mirrors were treated as single-sided, with the back-side coatings of the thin plates being neglected; however, as will be described further on, these reduce device efficiency by less than $\SI{10}{\percent}$. The mirrors were inserted into grooves machined into two aluminum base plates of a casing, defining the required elliptic paths with semi-minor axes ranging from $b_{0} \simeq \SI{17}{mm}$ to $b_{7} \simeq \SI{4}{\milli\meter}$. The divergence of the incident beam delivered by the crystal monochromator of MIRA could be covered with mirrors installed only on one side of the optical axis, as visible in Fig.\,\ref{fig:real_optic}. The casing was placed between two iron plates connected with NdFeB magnets, forming a yoke with a field of $\approx \SI{50}{\milli\tesla}$, sufficient to magnetize the FeSi supermirrors.\\

\begin{figure}[htb]
\centering
\includegraphics[width = \linewidth]{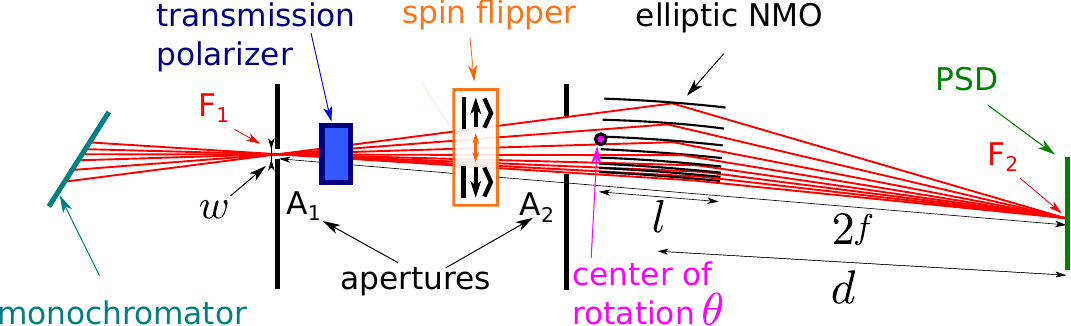}
\caption{Geometry of the experimental setup installed at the instrument MIRA at FRM II (top-down view). Studies with unpolarized neutrons were performed using the beam-defining apertures A$_1$ and A$_2$, the NMO prototype and a position-sensitive detector (PSD). For polarized neutrons, a transmission polarizer and a spin flipper were installed together with suitable magnetic fields to guide the neutron polarization between the polarizer and the NMO. (See text for more details.)}
\label{fig:sketch_geometry}
\end{figure}

The experimental setup for the investigation of the neutron optical properties of the NMO prototype is depicted in Fig.\,\ref{fig:sketch_geometry}. The monochromator of MIRA provides neutrons with a wavelength of $\lambda = \SI{4.9}{\angstrom}$, with $\rm{\delta}\lambda/\lambda \approx\SI{1}{\percent}$. A slit (A$_1$) with remotely-adjustable width $w$ defines a neutron source at the first focal point F$_1$, which the NMO is expected to image onto its second focal point F$_2$. A second slit (A$_2$) restricts the beam width at the entrance of the NMO to its geometric acceptance. This beam preparation allows us to define the neutron transport efficiency of the NMO as the ratio of two integrated neutron rates: that encompassing the focused beam image and that comprising all neutrons arriving at the detector with the NMO removed.\\

The NMO was mounted on a rotation table with its vertical axis of rotation centered in the entrance area of the NMO. This ensured that the exposed area of the NMO did not change upon rotation within the small range of angles $\theta$ needed to scan through the optimum of beam focusing. The neutrons were counted by a position-sensitive detector (PSD), installed at a distance $d$ from the center of the NMO (corresponding to $z=0$ in Fig.\,\ref{fig:nestedsketch}). Its large area of $\SI[product-units=power]{200 x 200}{\mm}$ and spatial resolution (FWHM) of $\SI{2.5\pm 0.1}{\milli\meter}$ \cite{kohliCASCADEMultilayerBoron102016} not only allowed us to resolve the complete beam into distinct reflected and transmitted components, but also to detect spurious partial beams due to misalignment of neutron optical components during the installation and alignment process. In addition, the PSD was mounted on a motorized bench, making it possible to vary the distance $d$, and thereby investigate the focusing properties of the NMO around F$_2$ along the optical axis.\\

For measurements with polarized neutrons, the beam was prepared by inserting a transmission polarizer behind A$_1$. This device consisted of a stack of silicon plates coated with polarizing supermirror, and a collimator to remove the reflected beam with the unwanted polarization state. Notably, the installation of the transmission polarizer does not alter the course of the transmitted and polarized beam from that previously defined in the setup without polarization. A flat-coil spin flipper allowed us to invert the polarization of the beam incident on the NMO with respect to its static magnetic field. In the limited space around the beam, a number of magnetic coils were installed between the polarizer and the NMO. These were tuned in such a way as to respect the field requirements of the spin flipper and to ensure a smooth magnetic guide field. All adjustments were done in pursuit of an optimum ``flipping ratio'', defined as the ratio of count rates in the focused beam image with the flipper switched on and off, respectively. In its final configuration, with a guide field exceeding $\SI{20}{G}$ everywhere between the polarizer and the NMO, we attained a flipping ratio of 16 and deemed this sufficient to study the polarization dependence of the various reflected and transmitted beam components behind the NMO.\\


\section{Measurements and experimental results}
\label{sec:meas_res}

\begin{figure}[htb]
\centering
\includegraphics[width = \linewidth]{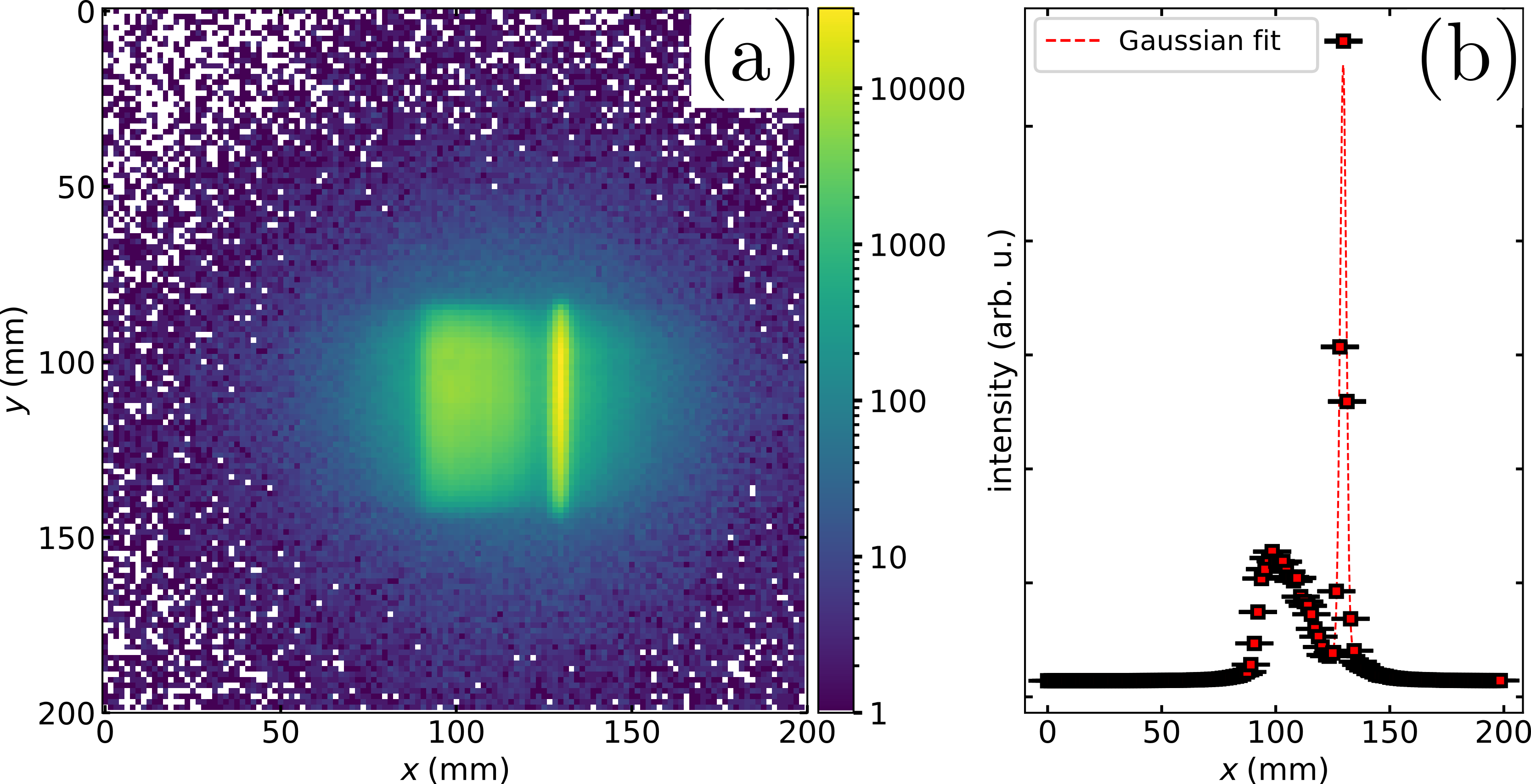}
\caption{Detector image of the measured intensity on a logarithmic scale (a), and after vertical integration on a linear scale (b), for an unpolarized beam of width $w=\SI{0.5}{\milli\meter}$. The orientation of the NMO and the distance to the detector are $\theta=0$ and $d=\SI{600}{\milli\meter}$, respectively, i.e., their optimal values. A sharp peak of focused ``spin up''-neutrons occurs to the right of a broader intensity distribution, whose large spread is mostly attributable to non-reflected ``spin down''-neutrons.}
\label{fig:best_focus}
\end{figure}

A first set of measurements was performed using the unpolarized beam. A global maximum in the detected focused intensity was found by independently varying two parameters: the orientation of the mirror device, $\theta$, and the distance between the NMO and the detector $d$. The amplitude of the peak of the vertically-integrated (i.e., along the $y$-axis of Fig.\,\ref{fig:best_focus}) detector data was chosen as a metric for the quality of focusing. For every combination of $\theta$ and $d$ on a grid, a Gaussian-type function was fitted to this peak in order to determine its horizontal position, $x_0$, amplitude, $A$, and width, $\sigma$, in addition to a constant background, $C$, i.e.,
\begin{equation}
\label{eq:gaussian}
I(x) = A \exp{\left(-\frac{(x-x_0)^2}{2\sigma^2}\right)} + C.
\end{equation}

\begin{figure}[htb]
\centering
\includegraphics[width = \linewidth]{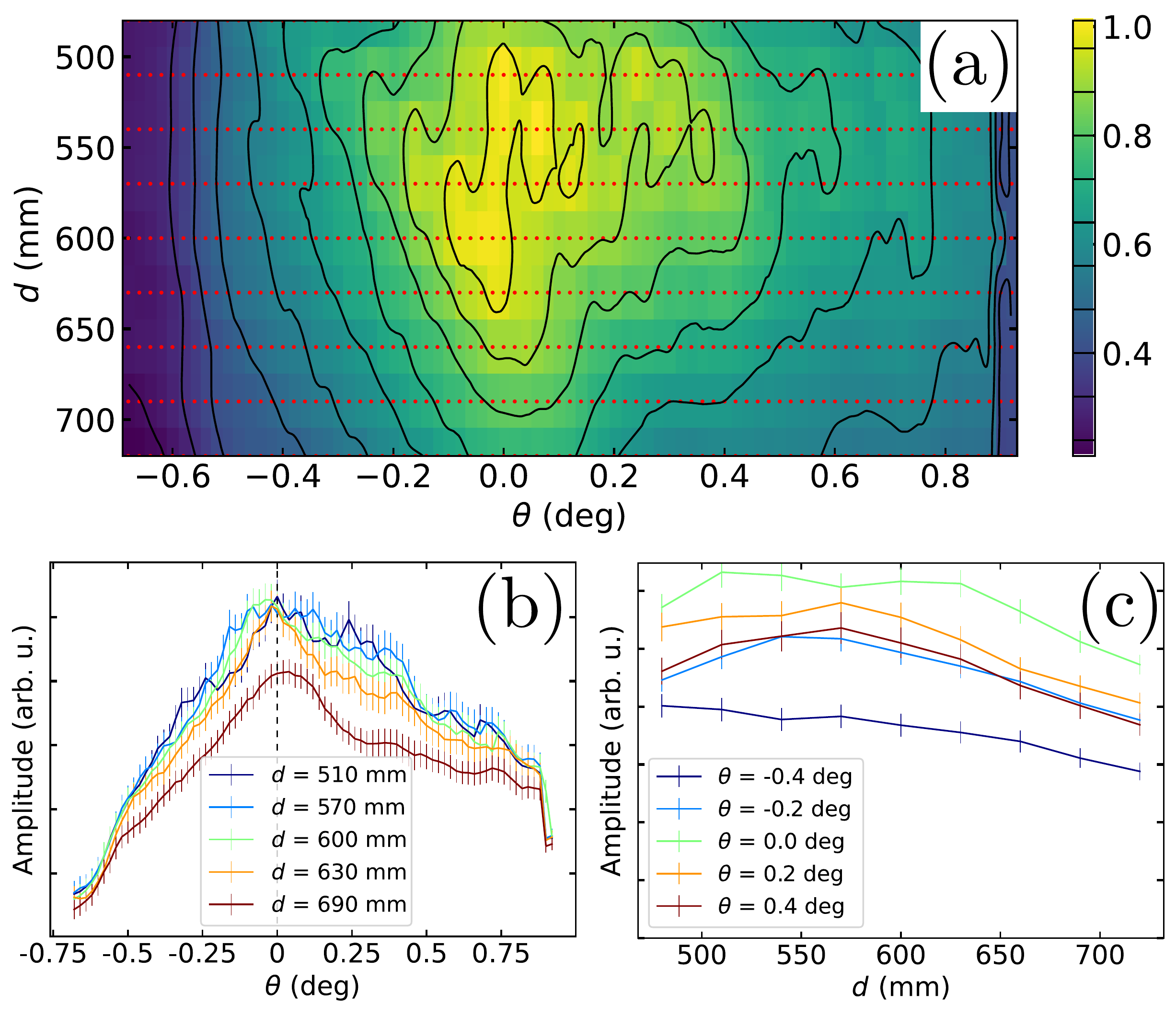}
\caption{a): Contour plot of the amplitudes $A$ for the unpolarized beam setup; data were acquired for the combinations of $\theta$ and $d$ indicated by red dots. The maximum visible at $\theta = 0^\circ$ is broadened due to the limited resolution of the detector. b) and c): Cuts through the contour map at fixed values of $d$ and $\theta$, respectively. The error bars include, besides counting statistical fluctuations, a fitting error of the non-Gaussian peak shape of both, the intensity of focused neutrons and the point spread function of the employed PSD \cite{kohliEfficiencySpatialResolution2016}.} 
\label{fig:orientation_heatmap}
\end{figure}

Figure \ref{fig:orientation_heatmap} a) shows a contour plot of the fitted amplitudes $A$ for all $\theta$ and $d$, in addition to cuts through this plot at constant $d$ (Fig.\,\ref{fig:orientation_heatmap} b)) and $\theta$ (Fig.\,\ref{fig:orientation_heatmap} c)), respectively. The global maximum was used to define the zero of $\theta$. The intensity peaks within a fraction of a degree about $\theta = 0$. The distance $d$ is a less critical parameter for optimal focusing; within uncertainties, the amplitudes stay constant within $d = \SI{600 \pm 30}{\milli\meter}$. These observations are consistent with expectations for this setup.\\

\begin{figure}[htb]
\centering
\includegraphics[width = \linewidth]{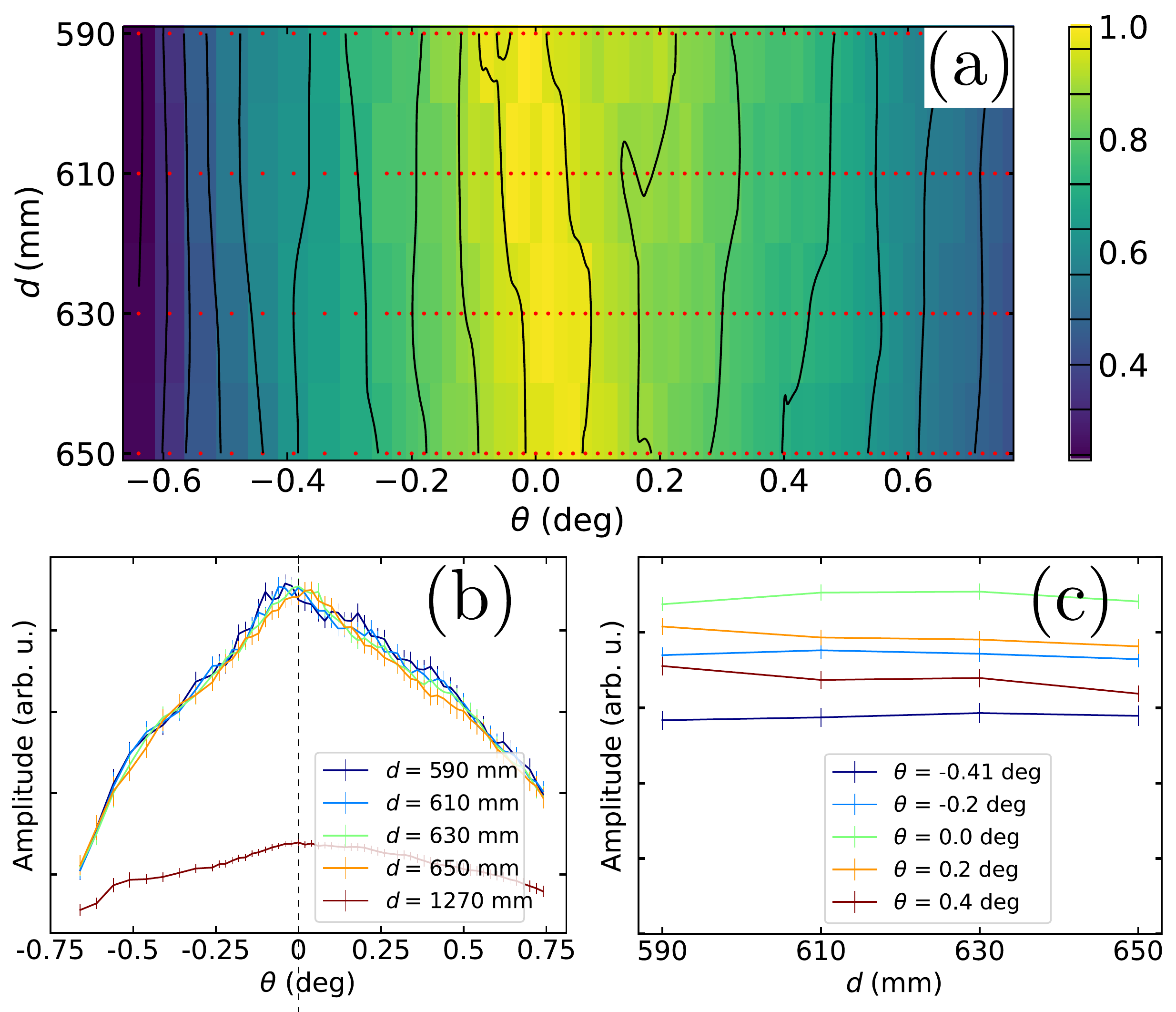}
\caption{a) Contour plot of the amplitudes $A$ for the polarized beam setup; data were acquired for the combinations of $\theta$ and $d$ indicated by red dots. The distribution with a distinct maximum at $\theta = 0^\circ$ is qualitatively equivalent to the results obtained for the unpolarized beam setup shown in Fig.\,\ref{fig:orientation_heatmap} (note the narrower range of $d$ values). Within uncertainties, the amplitudes $A$ measured at detector positions $\SI{590}{\milli\meter} \le d \le \SI{650}{\milli\meter}$ are indistinguishable. b) and c) show cuts through the contour map at fixed values of $d$ and $\theta$, respectively.}
\label{fig:second_night_map}
\end{figure}

Next, we implemented the components for beam polarization as described in the previous section. A repetition of the parameter optimization produced results similar to those obtained with the unpolarized beam (compare Fig.\,\ref{fig:second_night_map} to Fig.\,\ref{fig:orientation_heatmap}). Note that, in all studies reported, the finite resolution of the detector limited the study of sharp features in the intensity distribution.\\

After fixing $\theta$ and $d$ at values corresponding to optimum focusing, we varied the aperture A$_1$ to investigate the efficiency of neutron transport for several beam widths in the range $\SI{0.25}{\milli\meter} \leq w \leq \SI{6}{\milli\meter}$. To ensure that none of the neutrons incident on the NMO missed the device, we set the width of aperture A$_2$ to $\SI{8}{\milli\meter}$. Analysis of the detected intensities showed that, as a result of this measure, only five of the available eight elliptic mirrors of the prototype were illuminated.\\

\begin{figure}[htb]
\centering
\includegraphics[width = \linewidth]{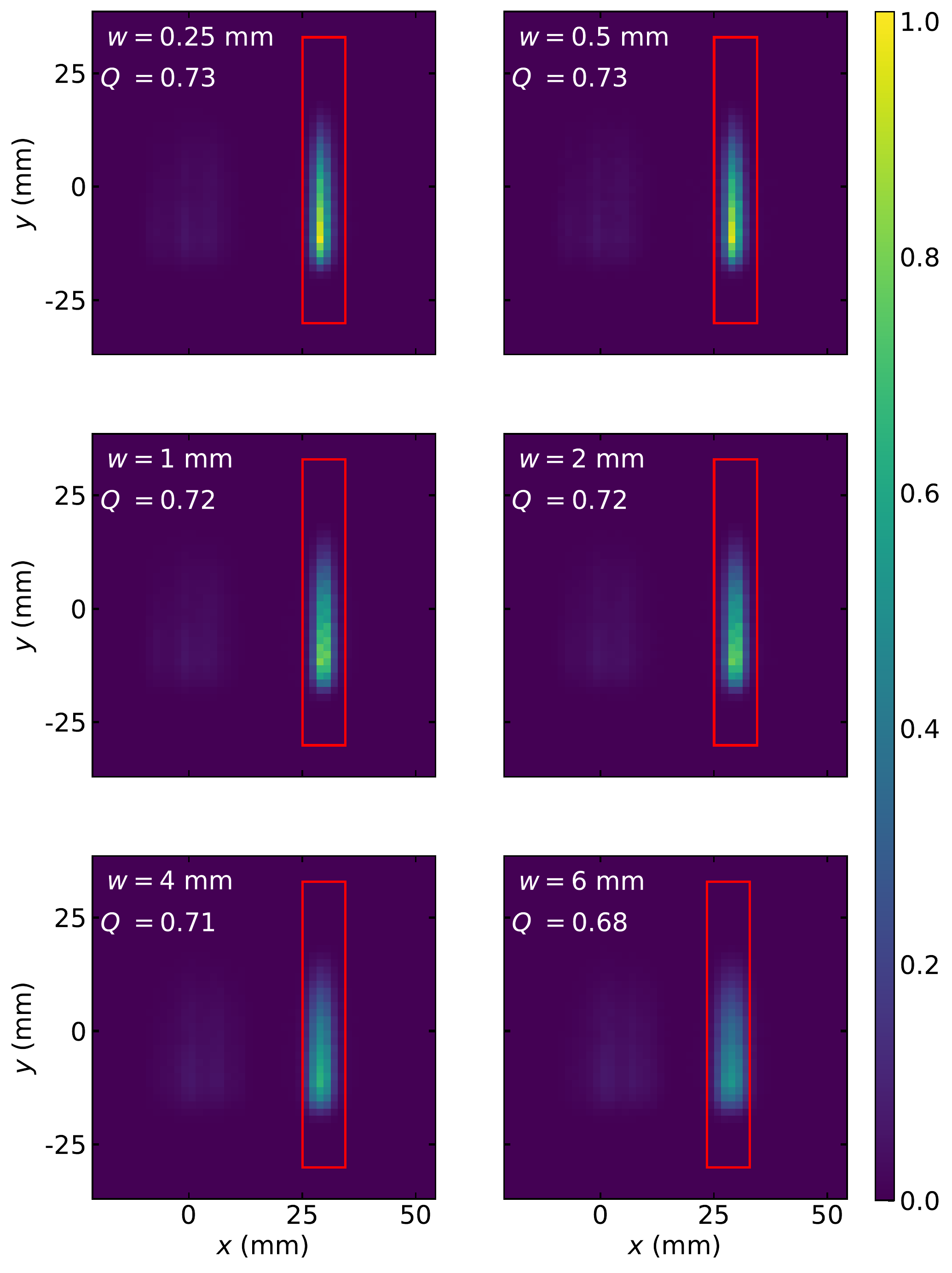}
\caption{Detector images used in the determination of the experimental figure of merit $Q(w,\, w_{\text{int}})$ defined in Eq.\,\ref{eq:FoM}, for various beam widths $w$ at F$_1$. The area of integration with width $w_{\text{int}}=\SI{9}{\milli\meter}$ and height $h = \SI{62.5}{\milli\meter}$ for the determination of $I_{F_2}$ is outlined by a red rectangle. The neutrons leaking through the NMO are barely visible in a diffuse region left of the intense peak of focused neutrons. The transport efficiencies amount to $Q\approx\SI{72}{\percent}$ for most $w$ with a slight drop at the largest width, as expected due to geometrical neutron losses. All images are normalized to their total intensity.}
\label{fig:efficiency_all}
\end{figure}

The corresponding normalized detector images are shown in Fig.\,\ref{fig:efficiency_all}. They all share a common structure, displaying a peak of focused neutrons and, barely visible left of this peak, an extended region of unfocused neutrons leaking through the NMO mirrors. With increasing $w$, one observes an increase in the width of the focal spot, as well as an increase in neutron leakage through the NMO, in agreement with the theoretical expectation.\\

For each $w$, we determined a rate $I_{F_2}$ of neutrons integrated over a $w_{\text{int}}=\SI{9}{\milli\meter}$ wide window (shown as red boxes in Fig.\,\ref{fig:efficiency_all}), which contains the whole focused intensity for the widest $w$. A second rate $I_{A_2}$ of the primary beam was obtained by counting all neutrons arriving at the detector with the NMO removed from the setup while keeping all other parameters constant. To characterize the efficiency of the imaging neutron transport, we then defined the figure of merit,
\begin{equation}
\label{eq:FoM}
Q(w,\, w_{\text{int}}) \coloneqq I_{\mathrm{F_2}}(w,\, w_{\text{int}})/I_{\mathrm{A_2}}(w).
\end{equation}

The obtained $Q$-values, also shown in Fig.\,\ref{fig:efficiency_all}, saturate for small widths $w$ at $Q\approx\SI{73}{\percent}$. The observed decrease of efficiency with increasing width results from an increasing fraction of neutrons that either pass the device unreflected or become reflected twice in one of its mirror channels. These purely geometric effects are inherent to NMO and persist even for perfect geometry. They scale with the ratio of $w$ to the ellipse's semi-minor axis, and, hence, become smaller with increasing lateral size of the NMO \cite{zimmerMultimirrorImagingOptics2016}. This and other size-related effects will be further discussed in Section \ref{sec:ellipticNMO}.\\

We note that a key quantity to be considered in the design of a neutron transport system is the ``brilliance transfer'' \cite{klenoSystematicPerformanceStudy2012}, which characterizes the efficiency of neutron transport from a source to a target area in an infinitesimal volume element of beam phase space. In practice, one has to optimize the brilliance transfer within an extended phase space volume, which is determined by the intended use of the beam. Both a toroidal and double-planar NMO cover a solid angle dictated by NMO geometry, and make use of a range of neutron wavelengths that is limited by the $m$-value of the supermirrors. One may therefore define an integrated brilliance transfer from a spot F$_1$ to an equally-sized spot F$_2$ to be the efficiency of neutron transport within the angular and wavelength acceptance of the NMO.\\ 

For our single-planar NMO, which refocuses neutrons only in one dimension, it is useful to define a partly-integrated brilliance transfer in the same way as the figure of merit $Q$ defined in Eq.\,\ref{eq:FoM}, but for an integration width equal to the beam width, $w_{\text{int}}=w$,
\begin{equation}
\label{eq:brillianceTransfer}
B(w) \coloneqq Q(w,\, w).
\end{equation}
The reason for having considered a figure of merit $Q(w,\, w_{\text{int}})$ with $w_{\text{int}} > w$ for characterization of the refocusing efficiency of the NMO was to eliminate the influence of the limited detector resolution of $\SI{2.5\pm 0.1}{\milli\meter}$, which did not allow us to measure the true integrated brilliance transfer for small $w$. For the largest $w$ studied, however, one finds $B(\SI{6}{\milli\meter})=0.62$ without any correction for the detector resolution, close to $Q(\SI{6}{\milli\meter},\SI{9}{\milli\meter})=0.68$. It is to be noted that these experimental values already include neutron losses due to imperfect machining tolerances, imperfect neutron polarization and supermirror reflectivity, and neutron absorption in the Si-wafers and coatings.\\


\section{Comparison of experimental results with Monte-Carlo simulations}
\label{sect:comp_exp_sim}

To support the interpretation of the experimental results, we performed Monte-Carlo simulations, using the McStas software package \cite{willendrupMcStasIntroductionUse2020}. The components employed were derived from those used in \cite{khaykovichXrayTelescopesNeutron2011} but, unless otherwise specified, further include finite reflectivity, substrate thickness, absorption in silicon, and refraction at the silicon-air interface. For the simulated setup, modeled with mirrors whose geometry and reflectivity of the double-sided coatings are equivalent to those of their real counterparts, we obtain reasonable qualitative agreement with measured data, as listed in Table\,\ref{tab:efficiencies_main}. We attribute the bulk of the remaining difference to the detector resolution and to imperfect polarization. Estimates of the latter were not included in the simulations due to insufficient knowledge of the separate efficiencies of the polarizer, the NMO, and the spin-flipper. That a large part of the discrepancy between $Q_{\text{exp}}$ and $Q_{\text{sim}}$ may be attributed to imperfect polarization is supported by the following argument. If one assumes that the polarizer provides a beam of a high polarization of $p>\SI{98}{\percent}$ \cite{georgiiMultipurposeThreeaxisSpectrometer2018}, the observed flipping ratio of 16 would correspond to a polarization of $\SI{90}{\percent}$ behind the NMO, which would produce an even larger discrepancy between experimental and simulated $Q$-values than observed. We also expect that the non-simulated finite mirror waviness has a negligible influence on the disagreement between $Q_{\text{exp}}$ and $Q_{\text{sim}}$, due to the small distance between the NMO center and the detector (only $\SI{600}{\milli\meter}$).\\

\begin{table}[]
\centering
\begin{adjustbox}{width=\columnwidth}
\begin{tabular}{l|llllll}
$w$ (mm) 	& 0.25 & 0.5  & 1    & 2    & 4    & 6    \\ \hline
$Q_{\text{exp}}$ 	& 0.73 & 0.73 & 0.72 & 0.72 & 0.71 & 0.68 \\ \hline
$Q_{\text{sim}}$  & 0.82 & 0.81 & 0.81 & 0.80 & 0.77 & 0.73
\end{tabular}
\end{adjustbox}
\caption{Measured and simulated figures of merit, $Q$. The simulation reproduces the experimentally-observed plateau for widths $w \leq\SI{4}{\milli\meter}$. We attribute the higher level of $Q_{\text{sim}}$ to the assumption of ideal polarizing components.}
\label{tab:efficiencies_main}
\end{table}

\begin{figure}
\centering
\includegraphics[width = \linewidth]{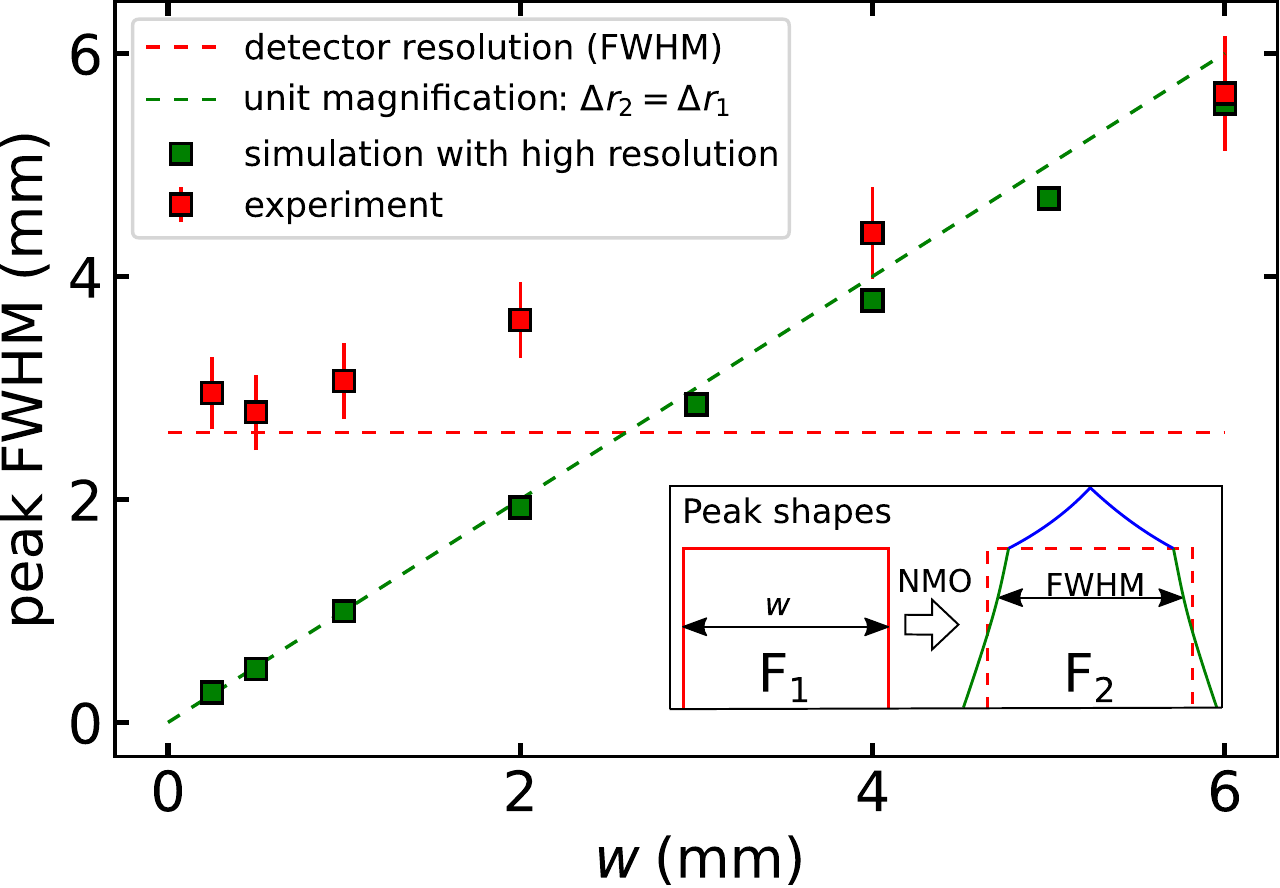}
\caption{Dependence of the horizontal width (FWHM) of the focused beam at F$_2$ on the width $w$ of the primary, rectangular beam at F$_1$. The McStas simulation (green squares) agrees well with the naive expectation of ideal one-to-one imaging, i.e., $\left|\Delta r_{1}\right| = \left|\Delta r_{2}\right|$ (green broken line), whereas the measured FWHM values were limited by the finite resolution of the detector of $\SI{2.5\pm0.1}{mm}$, as indicated by the horizontal, broken red line. The inset illustrates the effect of geometric aberrations on the image of the initially rectangular intensity distribution as described in the text.}
\label{fig:slitsize}
\end{figure}

In order to investigate the dependence of the width of the focused beam on the primary width $w$, we vertically integrated the detector images of Fig.\,\ref{fig:efficiency_all}, and determined the full widths at half maxima (FWHM) of the resulting intensity distributions. Figure \ref{fig:slitsize} shows the obtained FWHM plotted against $w$, together with the results of a simulation done with unphysically high resolution. While the measured FWHM at small $w$ reflect the physically-limited resolution of the PSD employed, for larger $w$, experimental and simulated FWHM values are in good agreement.\\

The small discrepancy between the simulated FWHM and $w$, observed at large values of $w$, arises from a slight distortion of the imaging at F$_2$, which is a consequence of geometric aberrations in elliptic guide shapes as described previously (Eq.\,\ref{eq:focdefoc} and Fig.\,\ref{fig:aberrations}). Neutrons reflected near the exit of the NMO produce a triangular feature on top of the peak profile (shown in blue in the inset to Fig.\,\ref{fig:slitsize}), while those reflected near the entrance smear out the sides of the primary rectangular distribution (shown in green). As may be seen in the inset to Fig.\,\ref{fig:slitsize}, these two effects conspire to reduce the FWHM at F$_2$ in comparison to $w$ at F$_1$.\\

\begin{figure}
\centering
\includegraphics[width = 0.6\linewidth]{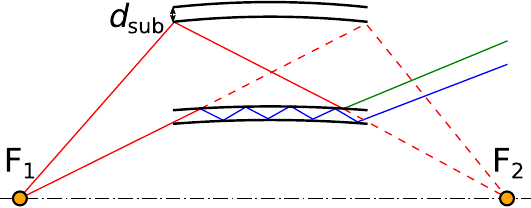}
\caption{Depiction of channeling (blue) and double reflections (green) for selected initial neutron trajectories (red) in an assembly of double-side coated mirrors, leading to a reduction of the figure of merit, $Q$. Dashed red lines show the unperturbed neutron trajectories for very thin mirrors.}
\label{fig:channeling}
\end{figure}

In addition to the simulations of the measurements presented in Section \ref{sec:meas_res}, we also performed simulations to study the influence of substrate thickness and mirror reflectivity on the figure of merit, $Q$. In these simulations, which are described in detail in \ref{app:simulation_optic}, the basic geometry of the NMO prototype, with its double-side coated mirrors, was maintained. The main result of these studies was that $Q$ was found to decrease with both increasing substrate thickness and beam width. Figure \ref{fig:channeling} depicts two effects that contribute to a reduction of NMO performance as a result of the finite substrate thickness. The first of these occurs when neutrons enter a substrate, either at the front surface of a mirror plate or due to leakage through the imperfectly reflecting inner mirror surface, and tend to become channeled by internal reflections. Although not necessarily absorbed within the substrate, such channeled neutrons are unlikely to reach the second focal point F$_2$. The second effect concerns those neutrons that are initially reflected close to the front end of one mirror, and then reflected again near the back end of the adjacent inner mirror. In future work beyond the current prototype, such effects may be avoided using custom single-side coated mirrors. (See \ref{app:simulation_optic} for an analysis of these design factors.)\\

\section{Applications}
\label{sect:applications}

This section discusses potential applications of NMO. Foremost among these is perhaps the possible use of NMO to solve the problem to extract neutrons efficiently from a small source, that is typically limited by ``phase space dilution'' associated with standard methods that use neutron guides. It is the opinion of the authors that such an application, in and of itself, is a good reason to explore the use of NMO in developing delivery systems for future neutron facilities.\\

The experiments reported above, with the small, elliptic prototype NMO, have demonstrated an efficient, one-dimensional refocusing of neutrons from a source to an instrument, mimicked by an aperture and a PSD, respectively. This shows that an NMO can serve as a complete neutron delivery system in a single, compact device. For an assessment of size-related effects, including losses due to gravitational bending of neutron trajectories, we report here results of additional McStas simulations for planar elliptic NMO.\\

Besides elliptic NMO, parabolic NMO offer interesting complementary functionality. Instead of refocusing the beam directly, the latter transform a high-divergence beam to low divergence, associatd with an increase in beam size. The original phase space volume as extracted from the moderator can be recovered using a second parabolic NMO that refocuses the neutrons. We will show that a system consisting of two parabolic NMO connected by a long guide can provide a high brilliance transfer over length scales of hundreds of meters.\\

In addition, the complementary properties of elliptic and parabolic NMO provide a versatile framework from which a wide variety of application-specific beam lines may be conceived. Certain components might even be integrated into existing guide systems at well-established research facilities to act as a final stage for neutron delivery, for example, in focusing a beam onto a small sample inside a pressure cell. Although such a hybrid approach would not necessarily offer the same brightness gains as would a fully NMO-based neutron extraction system, it might nevertheless achieve significant improvements in beam quality and a  reduction in background at the point of focus.\\


\subsection{Beam Extraction}
\label{sect:beam-extraction}

The extraction of neutron beams from an intense neutron source is, for various reasons, a challenging task. Some of the difficulties are purely technological. For instance, radiation damage occurring near the source requires that careful choice of materials be made, and further entails a need for regular replacement of neutron optical components in a highly activated area. Other challenges are more theoretical in nature: for example, how one is to maximize the efficiency of delivery of a neutron beam, characterized by its extent, divergence, and wavelength spectrum, to an instrument. This operation is fundamentally limited by Liouville's theorem, a corollary of which dictates that, for systems obeying Hamilton's equations of motion, the phase space density cannot increase. Optimally then, the efficiency would near the limiting value for the situation in which the density at the point of delivery is the same as that emitted at the surface of the source. NMO were conceived in order to approach this limit in neutron extraction even from a large solid angle off a moderator. As an added benefit, the implementation of NMO does not require that sensitive optical components be placed in close proximity to the source, and thereby mitigates some of the technological challenges described above.\\

\begin{figure}
\centering
\includegraphics[width = \linewidth]{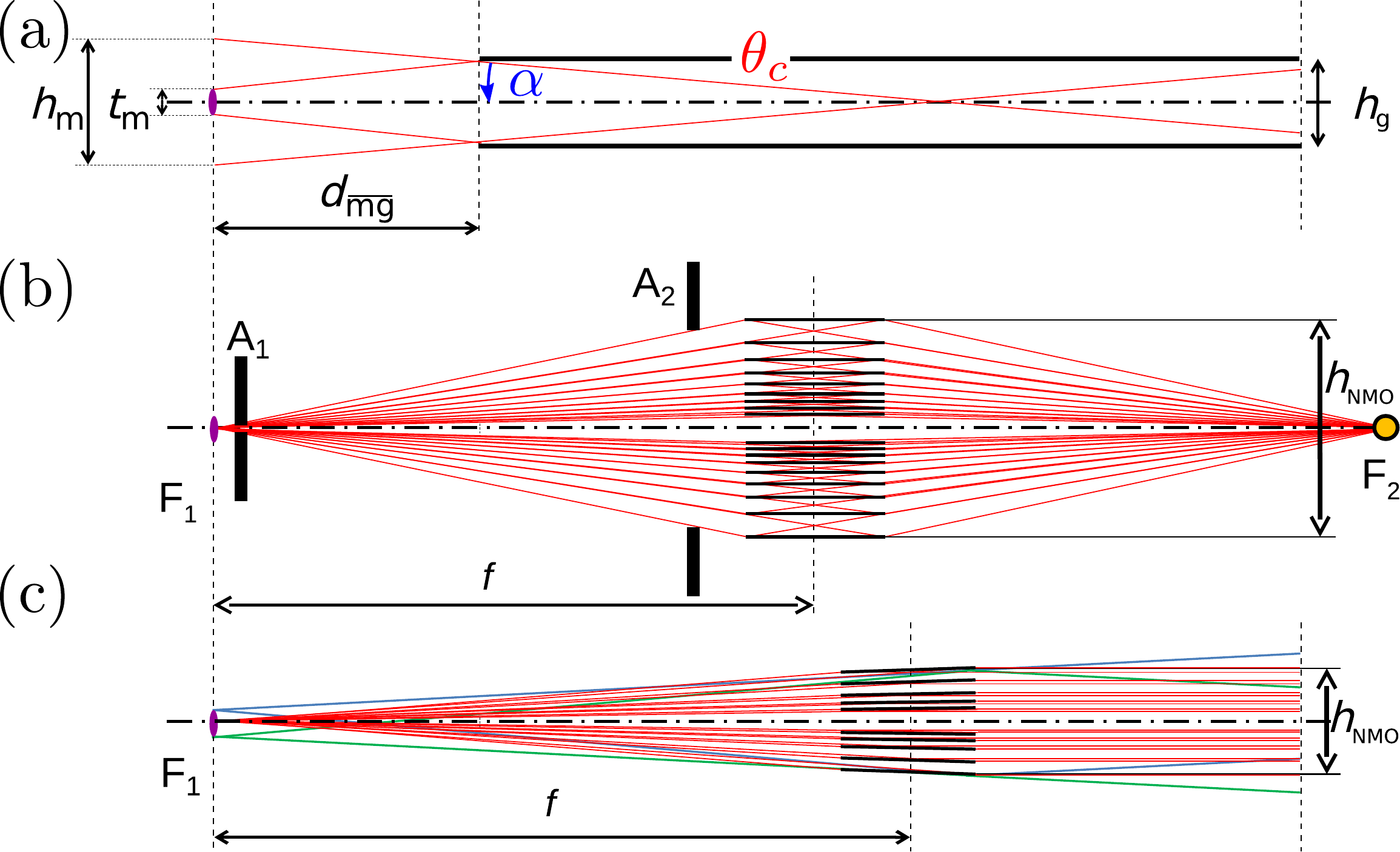}
\caption{Beam extraction from a moderator using a neutron guide (a), an elliptic NMO (b), or a parabolic NMO (c). Except for a central hole in solid angle, NMO can largely avoid the illumination losses occurring in neutron guides coupled to a small moderator. The elliptic NMO allows one to tailor the size and the divergence of the beam exactly to the needs, using apertures A$_1$ and A$_2$ far distant from the targeted focal point F$_2$. The parabolic NMO provides a low-divergence beam thus halving the angle of reflection compared to an elliptic NMO or a straight guide. }
\label{fig:beam_extraction}
\end{figure}

The standard method of neutron extraction employs neutron guides directed at the source. For technical reasons, guides are usually installed at a minimum distance $d_{\overline{\text{mg}}}$ from the radiating surface. As illustrated in Fig.\,\ref{fig:beam_extraction} (a), for a neutron guide with critical angle $\theta_{\text{c}, m}$ (Eq.\,\ref{eq:crit_angle}), the wavelength dependent ``vertical footprint'' of transportable neutrons at the position of the moderator is given by
\begin{equation}
h_{\text{m}} = h_{\text{g}} + 2d_{\overline{\text{mg}}} \tan\theta_{\text{c},m} \simeq h_{\text{g}} + 2d_{\overline{\text{mg}}} m \kappa \lambda.
\label{eq:footprint}
\end{equation}
For a complete illumination of the guide, the moderator height, $t_{\text{m}}$, must equal or exceed this footprint, i.e., $t_{\text{m}} \geq h_{\text{m}}$. According to Eq.\,\ref{eq:footprint}, this requires that the guide be narrower than the moderator, i.e., $h_{\text{g}} < t_{\text{m}}$. Otherwise, the incident phase space is diluted along the guide, reducing the brilliance at the sample position compared to the possible maximum determined by Liouville's Theorem. However, even for $h_{\text{g}} < t_{\text{m}}$, the guide will stay under-illuminated for wavelengths exceeding a cutoff $\lambda_{\text{c}}$ because the beam divergence is geometrically limited; the guide fails to extract the more divergent neutrons that it would otherwise be able to transport based on its angular acceptance. For $h_{\text{g}} > t_{\text{m}}$ on the other hand, the guide is under-illuminated for all wavelengths, lacking those neutrons with divergence larger than the angular acceptance of the extraction system, and additionally suffering from a dilution of phase space density within the acceptance.
\noindent A common definition of the extraction efficiency,
\begin{equation}
E_{\text{eff}} = \begin{cases}
t_{\text{m}}/h_{\text{m}} & \text{for}\ t_{\text{m}} < h_{\text{m}} \\
1 & \text{for}\ t_{\text{m}} > h_{\text{m}} ,
\end{cases}
\label{eq:extractionEfficiency}
\end{equation}
accounts for both types of guide under-illumination. Note that a further dilution of the beam phase space density occurs if the target area (with size smaller than $h_{\text{g}}$) is too far away from the guide exit.\\

Under-illumination of guides for neutron extraction seems to be an obstacle in the current trend toward small ``high-brilliance'' moderators, such as the flat, ``pancake'' shaped para-hydrogen moderator at the ESS \cite{Andersen2018, Zanini2019} or ``finger moderators'' \cite{cronert2016} for compact accelerator-based neutron sources \cite{carpenterDevelopmentCompactNeutron2019}. At the ESS, neutron beam optics assemblies (NBOAs) start $d_{\overline{\text{mg}}} \approx \SI{2}{\meter}$ \cite{schonfeldt2013} away from a flat moderator of thickness $t_{\text{m}} = \SI{30}{\milli\meter}$, whereas the extraction guides are all taller, i.e., $h_{\text{g}} > t_{\text{m}}$. Table \ref{tab:footprint} shows that, for typical NBOA parameters at the ESS, even for wavelengths as short as $\lambda = \SI{1}{\angstrom}$, the guide stays under-illuminated, as argued above: already at beam extraction, nearly half of the possibly useful neutrons are lost. For larger wavelengths and correspondingly increasing angular acceptance, these losses progressively increase.\\

As the authors of Ref.\,\cite{Andersen2018} state, ``The beam extraction efficiency suffers when the source is reduced to a size similar or smaller than the opening of the neutron guide. This results in a trade-off when reducing the source size, between the resultant brightness increase and the loss of beam extraction efficiency''. In full agreement with this statement, one notes that under-illumination of neutron guides entails two distinct types of compromises. On the moderator side, it hampers a free optimization of its dimensions for highest brilliance, which, in principle, is ideal for the investigation of small samples, thus conforming to another trend in neutron science highlighted in Section \ref{sect:introduction}. On the instrumental side, it restricts the possible scope of the instrument. Lacking the more divergent neutrons, which the guide would be able to transport if it were coupled to a larger moderator, a high-resolution instrument cannot simply be extended with an otherwise convenient high-intensity option.\\

The trade-off mentioned in the citation from Ref.\,\cite{Andersen2018} does not exist for NMO. The divergence accepted by an NMO is determined by the ratio of its height, $h_{\text{NMO}}$ (see Fig.\,\ref{fig:beam_extraction} (b) and (c)) to its distance from the moderator, $f$. Provided that a sufficient solid angle can be made available for neutron extraction, the transport of large-wavelength neutrons to an instrument needs no longer be accompanied by large losses of efficiency, as those quoted in Table \ref{tab:footprint}. Moreover, and as is further discussed in Section \ref{sec:ellipticNMO}, the size of an NMO that efficiently extracts neutrons from a moderator can be scaled down for decreasing moderator size, without loss of efficiency. These properties of NMO are an invitation to design high-brilliance sources that would allow for fully optimized neutron delivery to even the smallest of samples, without the compromises imposed by traditional neutron extraction systems.\\

\begin{table}[htb]

\centering
\begin{tabular}{c|c c c c c}
$\lambda$ ($\si{\angstrom}$) & $2\alpha\, (^\circ)$ & $h_{\text{m}}$ (mm) & $E_{\text{eff}}$ \\ \hline
0.5&     0.35   &  46.1   & 0.65 \\
1 &      0.69   &  57.7   & 0.52 \\ 
2 &      1.39   &  80.7   & 0.37 \\ 
4 &      2.78   &  127    & 0.24 \\ 
6 &      4.16   &  173    & 0.17 \\
10 &     6.93   &  265 	  & 0.11 \\
15 &     10.40  &  380	  & 0.08 \\ 
20 &     13.68  &  496	  & 0.06 \\
\end{tabular}
\caption{Vertical guide illumination losses of a typical NBOA at ESS, neglecting any further losses due to imperfect supermirror reflectivity and using the following parameters: $t_{\text{m}} = 30$ mm, $h_{\text{g}} = 34.6$ mm, $d_{\overline{\text{mg}}} = 1903$ mm, and $m = 3.5$. The guide-coating limited divergence is given by $2\alpha = 2\theta_{\text{c},m}$, and the efficiency of extraction of neutrons transportable in the guide is given by Eq.\,\ref{eq:extractionEfficiency}. Illumination losses in the horizontal direction further reduce efficiency, albeit to a less penalizing extent for a moderator with an assumed large width of $w_{\text{m}} = \SI{200}{\milli\meter}$.}
\label{tab:footprint}
\end{table}


\subsection{Brilliance transfer by a planar elliptic NMO}
\label{sec:ellipticNMO}

In this section we extend our analysis of a planar elliptic NMO with unit magnification as a basic solution for one-dimensional neutron extraction and refocusing transport to a targeted region near F$_2$. Here, our main goal is to study the influence of NMO size-dependent effects on the transport efficiency. Complying with the flat-moderator geometry discussed in the previous section, we consider here focusing in the vertical direction, which conveniently allows us to study the influence of gravity. We first recall several possibilities for spectral and angular beam definition, which NMO offer in addition to their superior beam extraction efficiency.\\

The geometry of a planar elliptic NMO defines the kinematics of imaging single neutron reflections. A neutron impinging on the $(n+1)$th mirror plate (see Fig.\,\ref{fig:nestedsketch} for the mirror indexing conventions) is reflected under a narrow range of angles centered around $\arctan{(b_n/f)}\approx {b_n/f}$, which becomes better-defined for smaller ratios $w/f$ and $l/f$. Since the angle of reflection for a certain wavelength is limited by the critical angle defined in Eq.\,\ref{eq:crit_angle}, the reflectivity edge of the supermirror produces a spectral cutoff for wavelengths shorter than $\lambda_{\text{c}, n}$, determined by the relation
\begin{equation}
b_n/f \approx m_n \kappa \lambda_{\text{c}, n}.
\label{eq:lambda_c}
\end{equation}
A common value $m_n = m$ for all mirrors thus results in a different $\lambda_{\text{c}, n}$ for each of the mirrors. However, since angles of reflection and mirror indices $n$ are correlated, one can generate a common cutoff for the whole covered angular range, $\lambda_{\text{c}, n} = \lambda_{\text{c}}$. According to Eq.\,\ref{eq:lambda_c}, this would require gradually decreasing $m$-values toward the center of the NMO. Implementation of spectrum shaping in this manner may render the use of auxiliary devices, usually employed to remove unwanted faster neutrons (e.g., Bragg filters), unnecessary, thereby avoiding any additional associated losses. In practice, selecting only a few $m$-values for producing a soft cutoff might be sufficient to prevent unwanted neutrons from reaching the instrument. A special option amenable to NMO would be the use of band-pass supermirrors to prepare a divergent monochromatic beam \cite{zimmerImagingNestedmirrorAssemblies2018}, similar to that in laterally-graded parabolic guides \cite{schneiderFocusingColdNeutrons2009}. Additionally, a beam could be polarized by using polarizing supermirrors in the construction of the NMO.\\

Concerning the angular beam definition, one can restrict the beam divergence by an aperture that masks a part of the mirrors. If the larger range is never needed, the NMO can be equipped accordingly with less mirrors. In the context of the reported experiments we had analyzed the imaging properties of a small, planar elliptic NMO with $N=8$ mirrors on one side of the optical axis. This ``half-device'' version can be extended to a ``full device'' with $N$ additional mirrors on the other side. For applications requiring a small beam divergence, such as, for example, in Small Angle Neutron Scattering, the half-device option provides a seamless coverage of the angular range $[\alpha_N \approx b_N/(f-l/2),\, \alpha_0 \approx b_{0}/(f+l/2)]$. The additional $N$ mirrors of a full device add the range of angles $[-\alpha_0,\, -\alpha_N]$. The total reflected beam thus contains a divergence hole, excluding the range $[-\alpha_N,\, \alpha_N]$ shown as $2\Delta\alpha$ in Fig.\,\ref{fig:nestedsketch}.\\

We now discuss simulations in the McStas software package of a full device, equipped with $2N$ identical single-sided $m = 4.1$ mirrors with edge reflectivity $R_0 = \SI{82}{\percent}$ and thickness $d_\text{sub} = \SI{0.15}{\milli\meter}$. To characterize its neutron transport efficiency, we determine the partly-integrated brilliance transfer $B$, as defined in Eq.\,\ref{eq:brillianceTransfer}. Note that we extend the integration over the whole angular range between the geometric extremes, $[-\alpha_0,\, \alpha_0]$, thus defining ``beam divergence'' in a conservative manner as an integral quantity that includes the divergence hole. For additional refocusing in the horizontal plane, one can use a double-planar device as described in Section \ref{sect:operationalPrinciple}, or, for a wide moderator and wavelengths that are not too large, a conventional ballistic channel of vertical mirrors, for which horizontal illumination losses might still be acceptable. Referring to linear independent components of neutron motion, the total integrated brilliance transfer in two dimensions is given by the product of the corresponding $B$-values per dimension, making it sufficient to simulate a single-planar elliptic NMO.\\

We are primarily interested in how $B$ changes with NMO size and neutron wavelength $\lambda$, for an NMO capable of transporting neutrons with the maximum possible divergence. To this end, we simulated the transport of a narrow range ($\pm 5 \si{\percent}$) of wavelengths around a central value $\lambda$, choosing for each $\lambda$ a corresponding $b_0$ so that the critical angle matches the geometrically-defined divergence $[-\alpha_0,\, \alpha_0]$. For a well-defined linear size scaling of the whole NMO, we then varied the focal length $f$ while keeping constant the ratios $b_0/f$ and $l/f$. For each situation, we implemented the maximum number of mirrors for the respective wavelength. Remembering that the distance between adjacent mirrors decreases toward the center of the NMO (see Fig.\,\ref{fig:nestedsketch}), the number $2N$ is limited by the condition that the minimum distance between mirror surfaces cannot be smaller than the sum of thicknesses of the substrate and the ridge between adjacent grooves, i.e., $d_{\text{min}}=d_{\text{sub}}+d_{\text{g}}$. For the latter we conservatively chose $d_{\text{g}} = \SI{0.35}{\milli\meter}$ as a technically reasonable lower limit. Taking $d_{\text{min}}$ as constant, $2N$ increases when scaling up the size of the NMO.\\

\begin{figure}
\centering
\includegraphics[width = \linewidth]{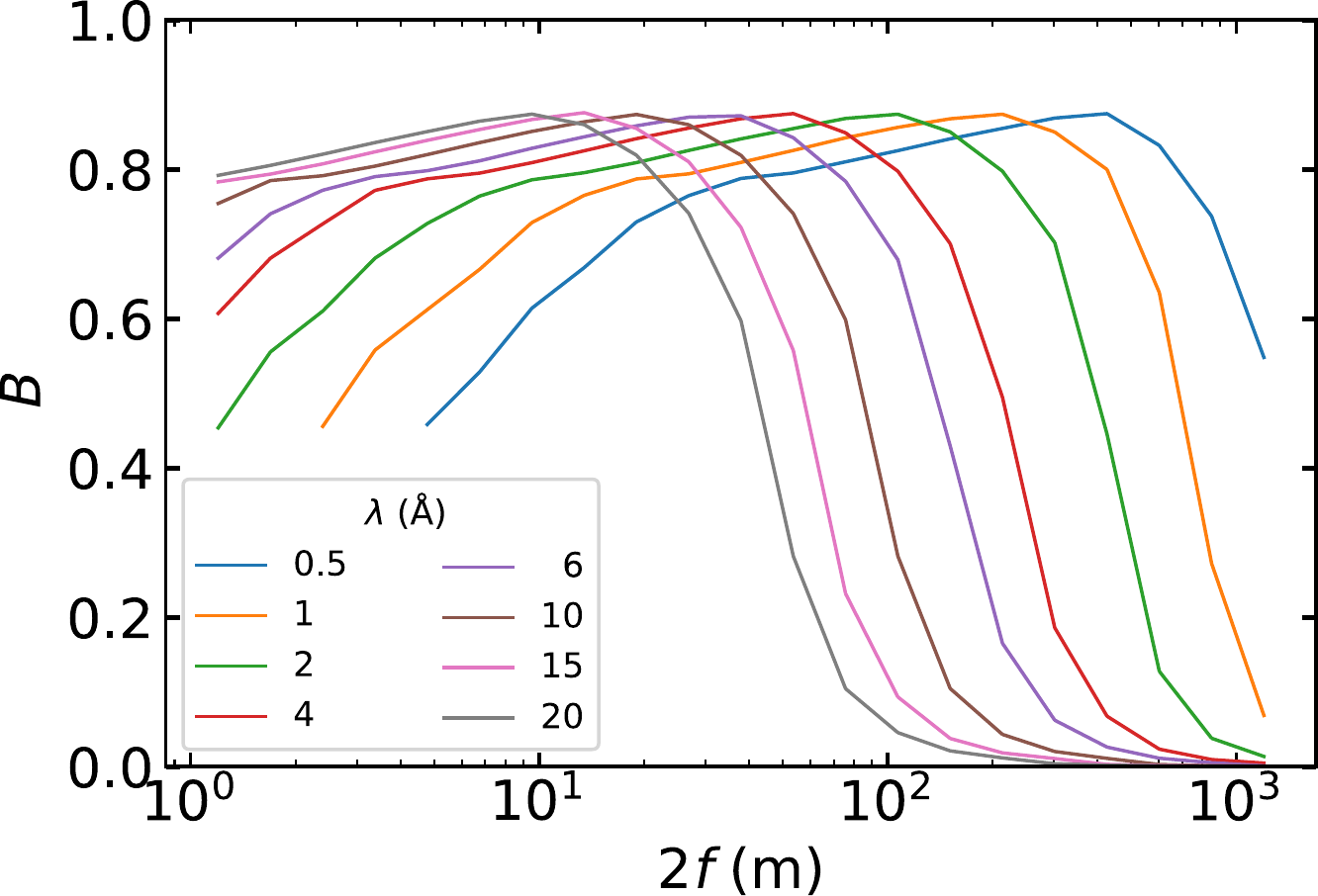}
\caption{Partially integrated brilliance transfer $B$, as defined in Eq.\,\ref{eq:brillianceTransfer}, for a planar elliptic NMO scaled in proportion to its focal length $f$ and assumptions as described in the text. The following parameters were kept constant: $l/f = 1/10$ , $d_{\text{sub}} = \SI{0.15}{\milli\meter}$, $m=4.1$, $R_0=0.82$, and $w = \SI{6}{\milli\meter}$.}\label{fig:NMO_scaling}
\end{figure}

The results shown in Fig.\,\ref{fig:NMO_scaling} can be understood as the superposition of several effects. First, one notes that, as a consequence of the adopted conditions, the larger one chooses $f$ (and thus $2N$), the smaller the resulting divergence hole becomes. A second effect favoring a larger NMO is the geometrical loss associated with neutrons being reflected twice or not at all when passing through one of the mirror channels. This loss scales with the ratio $w/b_n$ of the height of the beam window targeted at F$_2$ to a corresponding semi-minor axis of the ellipse. Gravity bends the neutron trajectories, violating the assumptions on which the recipe of NMO construction was based. While, in principle, one could adapt the mirror distances to account for gravity for monochromatic neutrons, we deemed this too restrictive for the present discussion. The dependence of gravity on $f$ is opposite to to aforementioned effects, finally limiting $B$ to an optimum value, which, for the assumptions made, stays well above $\SI{85}{\percent}$ within a broad range of values for $f$.\\

One can see from Fig.\,\ref{fig:NMO_scaling} that the $B$-values for the NMO are significantly larger than those for a representative NBOA at the ESS, quoted in Table \ref{tab:footprint}. Additional losses due to mirror reflectivity were solely taken into account for the simulated NMO and neglected for the NBOAs in Table \ref{tab:footprint}. However, the latter are expected to suffer greater losses as the number of required reflections is higher compared to an NMO. The divergence hole of the NMO may, at first, appear to be a disadvantage. However, it generates only a marginal loss of $B$, which is more than compensated for by a large gain in beam quality, especially considering that the beam spectrum transported by a neutron guide is unavoidably contaminated with fast neutrons.\\

Including other types of NMO, it is, for instance, possible to double the divergence per dimension with a setup comprised of two parabolic NMO in series, due to halving the angle of reflection compared to a single ellipse. In the next section we show that this idea is well-suited to neutron transport over a long distance, being much less limited by gravitational beam bending.\\

\subsection{Long-distance neutron transport}
\label{sec:transport neutrons}

As discussed in Section \ref{sec:ellipticNMO}, a well-designed elliptic NMO with unit magnification can, to a large extent, restore the beam phase space density at its second focal point, F$_2$, to that emitted at F$_1$. However, as may be seen in Fig.\,\ref{fig:NMO_scaling}, for focal lengths $f$ exceeding a few tens of meters, and for typical cold-neutron wavelengths, gravitational bending of neutron trajectories spoils the vertical brilliance transfer of a simple elliptic NMO. This problem also prevails in long focusing guides \cite{Weichselbaumer2015}, where it is responsible for significant neutron losses and phase space distortions. For instruments requiring a very long neutron flight path, such as for example a neutron time-of-flight (ToF) spectrometer with highest resolution, a different approach is needed.\\

\begin{figure}[htb]
\centering
\includegraphics[width = \linewidth]{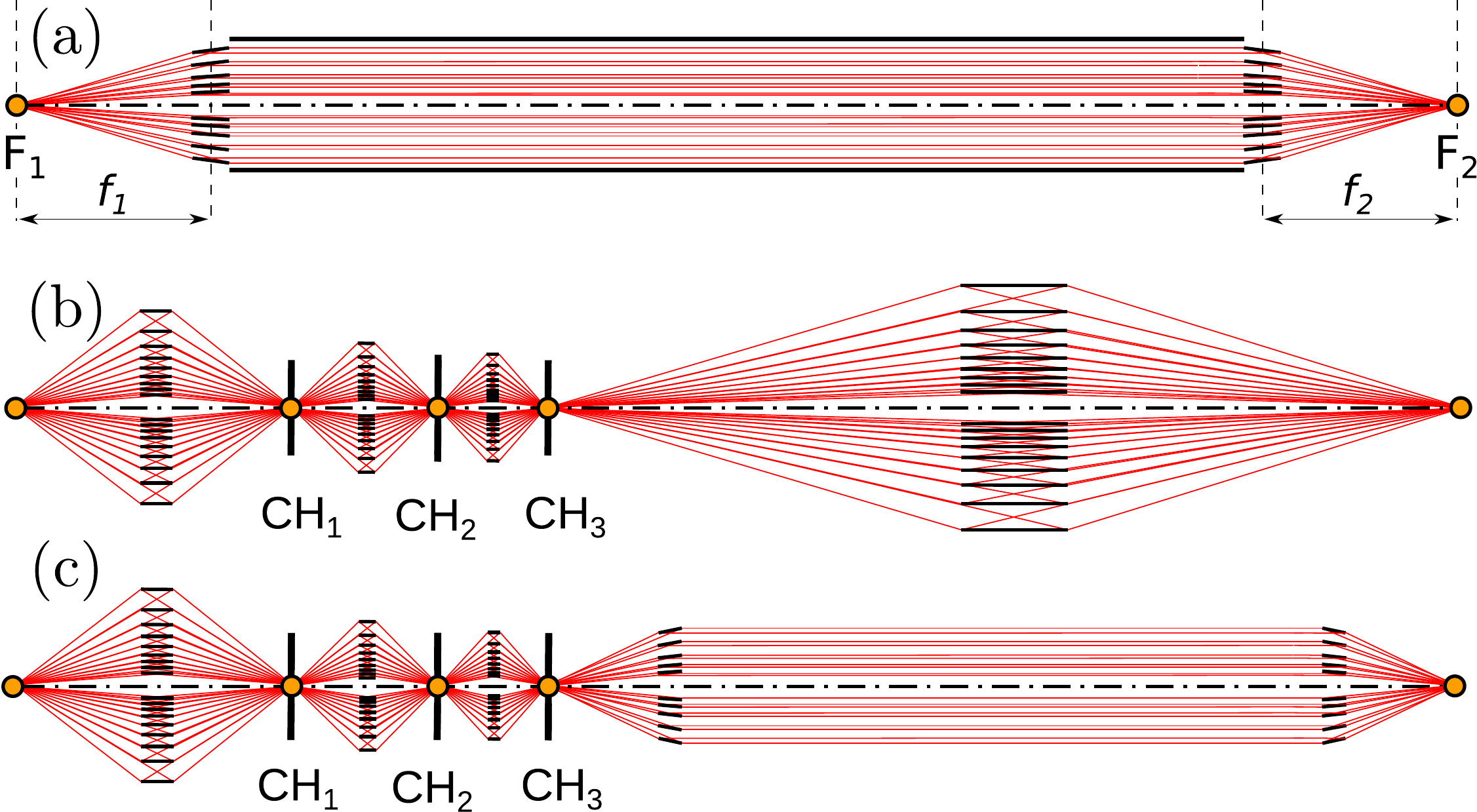}
\caption{Some options for long-distance transport of neutrons from a moderator to an instrument. (a) System of two parabolic NMO connected by a guide. Different focal lengths $f_1$ and $f_2$ may be used for non-unit magnification (see also Section \ref{sec:focusingMagnification}). (b) Elliptic NMO provide intermediate beam images for placement of choppers (CH$_1$ - CH$_3$). (c) Configuration with elliptic and parabolic NMO.}
\label{fig:Neutron-Beam-Transport}
\end{figure}

To cover such applications, we propose here a system consisting of two parabolic NMO connected by a long guide for low-loss transport of the produced low-divergence beam, as shown in Fig.\,\ref{fig:Neutron-Beam-Transport} (a). The focal lengths $f_1$ and $f_2$ can be configured according to the needs of the instrument. The first parabolic NMO, with its second focal point situated at infinity, transforms the divergent beam extracted from the moderator to low divergence, and increased beam size according to Liouville's theorem. This significantly reduces the number of reflections per meter in the guide, with a maximum reflection angle well below the critical angle. Therefore, reflection losses are small, so that the length of the guide only weakly affects the transported flux. The second NMO refocuses the low-divergence beam from the guide onto its second focal point.\\

\ref{app:pancake_extraction} presents simulations of such a system, comprised of two double-planar parabolic NMO in which a two-dimensional beam is extracted from a small, circular moderator with a diameter of $\SI{30}{\milli\meter}$ (which could be a ``finger moderator'' \cite{cronert2016}) and refocused with unit magnification after being transported over $\SI{160}{\meter}$ by a straight guide. The simulations show that gravitational effects only slightly disturb the transformed intensity distribution, such that a significant ratio of neutrons entering the first NMO is recovered at the focus. The integrated brilliance transfer thus determined for the whole size of the moderator amounts to $B_{\text{tot}}=\SI{23}{\percent}$.\\


Comparing such a system to an elliptic NMO, one notes, in addition to the advantages already mentioned, the following: i) the length of the straight guide section for long-distance transport can be chosen freely; ii) due to the small divergence of the beam provided by the first parabolic NMO, mirrors with small $m$-value can be used, taking advantage of a higher edge reflectivity at reduced cost; and iii) the lateral extent of the beam throughout transport stays smaller, reducing the amount of shielding necessary. Both systems have in common that only useful neutrons are transported. In order to avoid the direct line-of-sight from the experiment to the moderator, a vertically imaging NMO may be combined with a horizontal beam bender \cite{schaerpfPropertiesBeamBender1989, alefeldNeutronenleiterBerichtUber1965}, which channels neutrons within the empty space between curved mirror plates via multiple reflections.\\

Both elliptic and parabolic NMO can be combined for a versatile configuration of beam lines. If the installation of choppers in a beam line is required, elliptic NMO may be advantageous as they can refocus the beam onto the slit position of choppers (see Fig.\,\ref{fig:Neutron-Beam-Transport} (b)). If choppers and a long neutron flight path are required, as for high-resolution ToF instruments, a system of elliptic NMO followed by a pair of parabolic NMO connected by a guide (as discussed above) might be the best choice (see Fig.\,\ref{fig:Neutron-Beam-Transport} (c)).\\

\subsection{Focusing and magnification with NMO}
\label{sec:focusingMagnification}

\begin{figure}[htb]
\centering
\includegraphics[width = \linewidth]{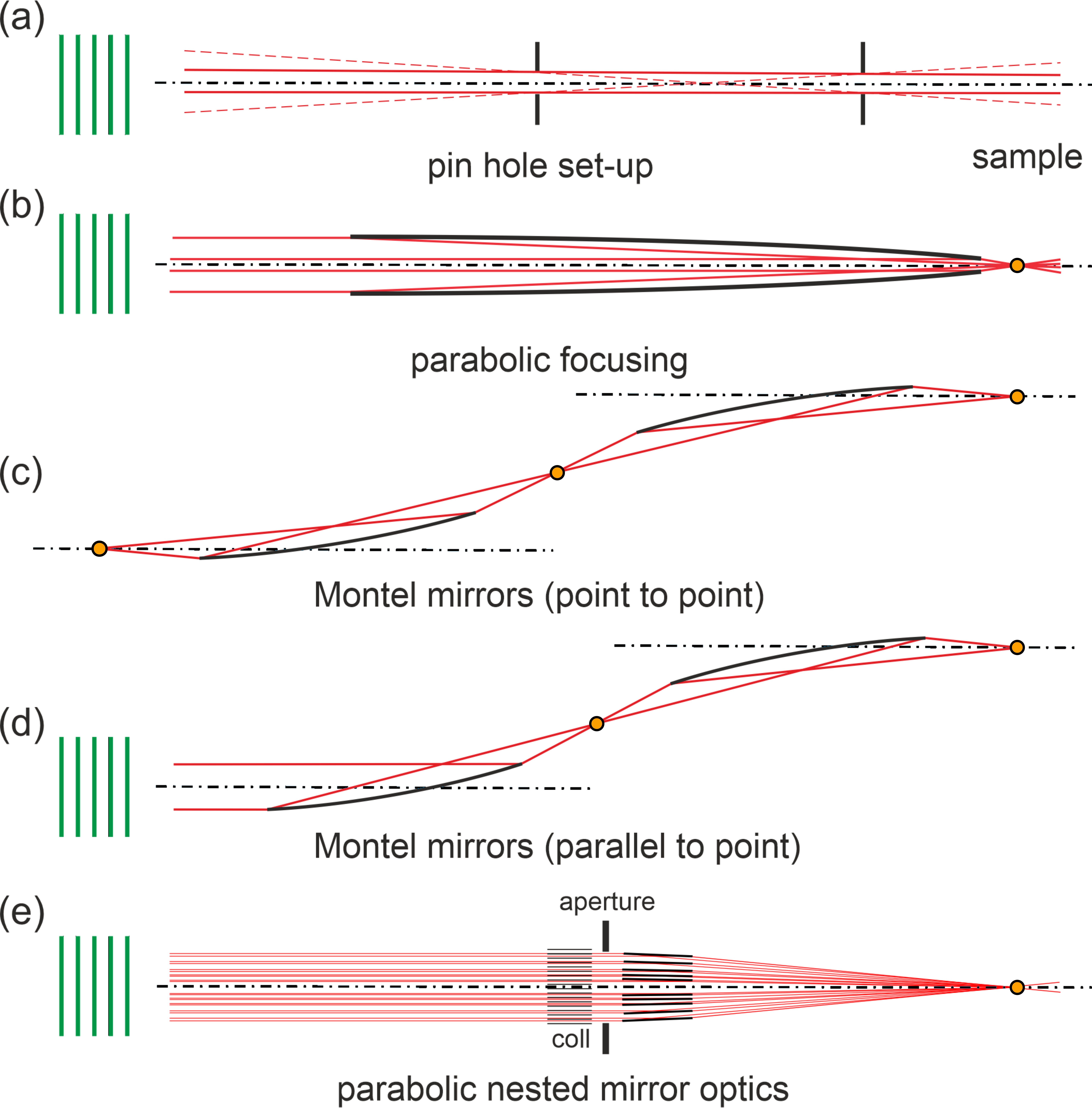}
\caption{Various options for illuminating a sample with a neutron beam (only one dimension is shown). 
An optimum beam definition at the sample requires slits (a) or focusing guides (b) placed close to the sample position, which is difficult to achieve for a bulky sample environment. A Selene set-up using parabolic (c) and/or elliptic (d) Montel mirrors requires significant space. Moreover they displace the beam from the original optical axis. Parabolic  NMO (e) provide a compact means for the beam definition at the sample.}
\label{fig:optics_20210528}
\end{figure}

Currently, neutron beams at the sample position are defined by means of collimators and apertures based on the principles of the pinhole camera (Fig.\,\ref{fig:optics_20210528} (a)). If the final aperture cannot be placed close to the sample, as, for example, if a bulky sample environment is used, sample surroundings will also be illuminated by a ``penumbra'' (broken red line), resulting in an increased background. Replacing the slits by elliptic or parabolic focusing guides \cite{boeni2014} as shown in Fig.\,\ref{fig:optics_20210528} (b) leads to a better definition of the beam at the sample, and an increase in flux density.\\

Still, if the guide exit is more than several tens of millimeters away from the sample, a diffuse ``halo'' develops around the focused beam. Moreover, in order to reduce phase space inhomogenities, the guide must be sufficiently long \cite{boeni2014}. If sufficient space along the beam is available, a ``Selene'' setup \cite{stahnFocusingNeutronReflectometry2016}, based on two elliptic Montel mirrors may be used, as shown in Fig.\, \ref{fig:optics_20210528} (c). However, to obtain a reasonable beam size ($\sim$ 10 mm), these systems become several meters long and are thus difficult to implement at existing beam lines. A partial setup is also not helpful: using only a single Montel mirror deflects the beam and strongly distorts its phase space \cite{Stahn2011}.\\

Most of the disadvantages of slit and guide geometries can be eliminated by using NMO. As demonstrated with the prototype, focal lengths of $\SI{600}{\milli\meter}$ can be easily achieved, thus allowing one to place the NMO approximately $\SI{500}{\milli\meter}$ from the sample, and providing plenty of space for the sample environment. Moreover, a parabolic NMO provides a beam with a large divergence, which, for this type of NMO, is given by four times the critical angle of reflection of the outermost supermirror. For example, an $m=6$ supermirror can generate a beam with a divergence of nearly $4.8^\circ$ for $\SI{2}{\angstrom}$ neutrons. For $m=8$, this is $1.6^\circ$ at $\lambda = \SI{0.5}{\angstrom}$. As shown in Fig.\,\ref{fig:optics_20210528}\,(e), the divergence and the size of the beam at the sample can be adjusted by an aperture and a collimator in front of the NMO \cite{Schanzer2018}.\\

As proposed by Stahn and Glavic, the performance of neutron reflectometers can be boosted by using a Selene setup of the types shown in Figs.\,\ref{fig:optics_20210528} (c) and (d) \cite{stahnFocusingNeutronReflectometry2016}. Typically, the neutron beam is extracted from the moderator by a parabolic Montel mirror, thus producing a focused beam, that is further transformed by a pair of elliptic Montel mirrors. Such a focusing system is capable of generating beams as small as 2 mm at the sample position, avoiding illumination of the surroundings.\\

In order to deliver a well-defined beam, geometric tolerances of the mirrors and gravity-induced distortions must be sufficiently small. Using sophisticated techniques for manufacturing, a waviness of $\eta \simeq 4\cdot10^{-5}$ rad can be achieved, leading to an estimated beam blurring on the order of $\Delta w \simeq 2\eta d_{\overline{\text{ot}}} = \SI{1.2}{\milli\meter}$ for an optics-target distance of approximately $d_{\overline{\text{ot}}} = \SI{15}{\meter}$ (Table \ref{tab:perfection-of-NMO}). The blurring can be reduced by decreasing $d_{\overline{\text{ot}}}$. However, a shortening of the mirrors will also reduce the beam size, $w \propto d_{\overline{\text{ot}}}$. For comparison, we also quote estimated parameters $\Delta w$ and $w$ for a short focusing guide \cite{hils2004}.\\

\begin{table}[htb]
\centering
\begin{adjustbox}{width=\columnwidth}
\begin{tabular}{c|c c c c c}
device  		& $d_{\overline{\text{ot}}}$  & $\eta$	   & $\Delta w$  & $w$ \\
			& (m)    	  & ($10^{-4}$ rad)  & (mm)  	 & (mm) \\ \hline
focusing guide \cite{hils2004}	& 0.5 		  & 1  		   & 0.1   	 & 1 \\ 
Selene 
 \cite{stahnFocusingNeutronReflectometry2016}  & 15  & 0.4  & 1.2   & 10 \\ 
NMO$_{\rm parab}$   		& 0.6 		  & 2  		   & 0.24 	 & 6  \\
NMO$_{\rm micro}$   		& 0.1 		  & 0.5  		   & $0.01$ 	 & 0.5  \\
\end{tabular}
\end{adjustbox}
\caption{Exemplary geometrical parameters of various focusing devices. The variables $d_{\overline{\text{ot}}}$, $\eta$, $\Delta w = 2\eta d_{\overline{\text{ot}}}$, and $w$ denote the distance of the optics to the target, the waviness, the blurring of the beam, and the beam size at the target, respectively. The values quoted for $\Delta w$ and $w$ are approximate, as they depend on beam divergence, neutron wavelength and the detailed design of the optical components.}
\label{tab:perfection-of-NMO}
\end{table}

Compared to Montel mirrors, NMO can deliver larger beams across shorter distances. For instance, the elliptic NMO used during our experiments delivered a beam with $w \simeq 6$ mm for $d_{\overline{\text{ot}}} = \SI{0.6}{\meter}$. A parabolic NMO$_{\rm parab}$ with a focal length of $\SI{0.6}{\meter}$ and an assumed waviness of $\eta = 2\cdot10^{-4}$ rad allows one to define the focused beam with a blurring on the order of only $\Delta w \approx \SI{0.24}{\milli\meter}$ (Table \ref{tab:perfection-of-NMO}). Further advantages of NMO for focusing are: i) the useful flux density at the sample is increased by a factor of four when compared to a Selene set-up due to the doubling of the divergence in each dimension, ii) the design of the beam line is more compact, iii) reflection losses are reduced as the number of reflections is halved and iv) the alignment of an NMO is as straightforward as for a single neutron guide element.\\

NMO may also offer an efficient means to focus neutron beams for prompt gamma activation analysis (PGAA). Because the NMO can be placed far away from the sample, neutron captures occurring in the NMO can be shielded very efficiently, thus reducing background.\\

\begin{figure}[htb]
\centering
\includegraphics[width = \linewidth]{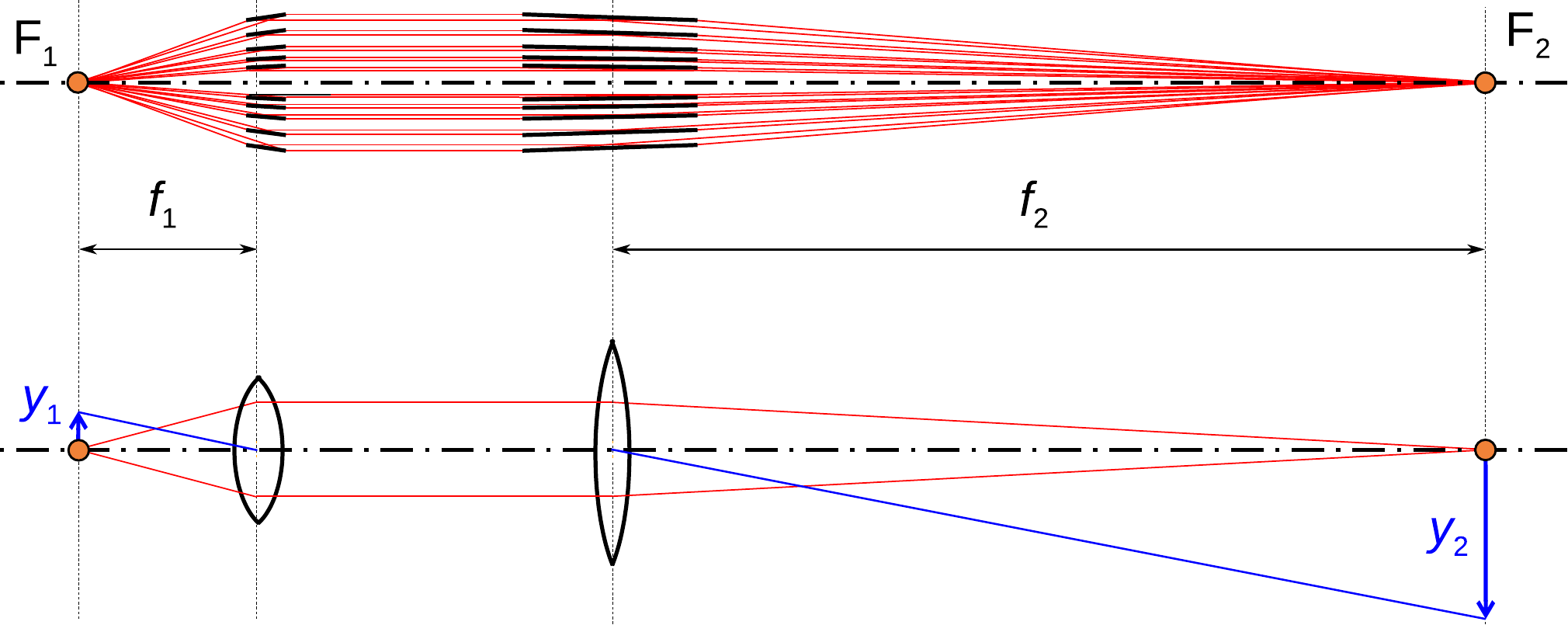}
\caption{Sketch of a lens system comprised of two parabolic NMO. The lower part of the figure shows an equivalent lens system for visible light optics. The magnification is given by $M = f_2/f_1$.}
\label{fig:lens_system}
\end{figure}

Combining two parabolic NMO with different focal lengths $f_1$ and $f_2$ allows for a magnification of neutron beams, see Fig.\,\ref{fig:lens_system}. Neutron lenses may be useful for imaging or microscopy with neutrons \cite{khaykovichXrayTelescopesNeutron2011, jorbaHighresolutionNeutronDepolarization2018}, to overcome the limited spatial resolution of current set-ups, which is of the order of $10\ \mu$m. The limitations are caused by the resolution of the neutron-sensitive scintillator and the available flux at the sample position \cite{Trtik2015}.\\


To achieve a resolution of $\SI{10}{\micro\meter}$ with an assumed waviness of $\eta = 5 \cdot 10^{-5}$ rad for the parabolic mirrors requires $d_{\overline{\text{ot}}} = \SI{0.1}{\meter}$ (NMO$_\text{micro}$ in Table \ref{tab:perfection-of-NMO})). In combination with the NMO$_{\rm parab}$, a magnification $M = 6$ would be achieved. However, for an NMO as small as NMO$_\text{micro}$, the distance between the reflecting mirror surfaces becomes of the order of $d_{\rm min} = \SI{10}{\micro\meter}$, i.e. much smaller than the typical thickness of Si wafers. Hence, the NMO may have to be assembled from stacked wafers of varying thicknesses, leading to prohibitively high manufacturing costs. Therefore, NMO may only be applicable for imaging and magnification purposes at intermediate length scales, i.e., on the order of several hundredths of millimeters.\\

\begin{figure}[htb]
\centering
\includegraphics[width = \linewidth]{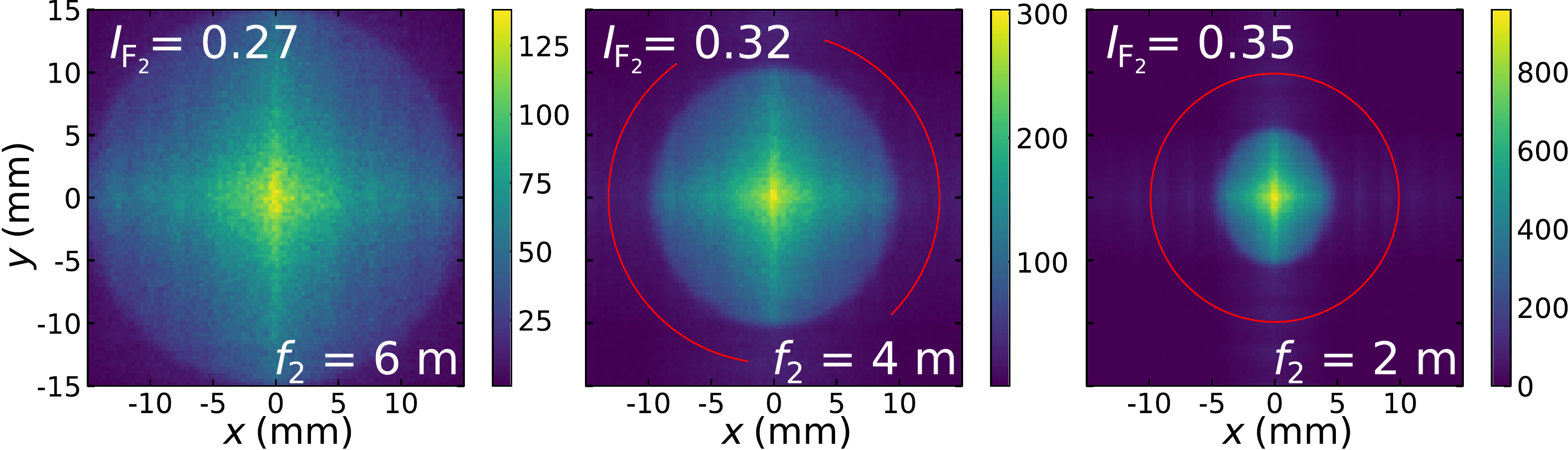}
\caption{Simulated intensity distributions at the sample position, F$_2$, of the neutron transport system as presented in \ref{app:pancake_extraction}. The focal length of the first NMO is kept constant $f_1 = \SI{6}{\meter}$, while the focal length of the second NMO, $f_2$, is varied between $\SI{6}{\meter}$ and $\SI{2}{\meter}$. The magnification is given by $M = f_2/f_1$. The fraction of neutrons arriving within the detector area at F$_2$ divided by the number of neutrons hitting the first NMO is given by $I_{\text{F}_2}$. Red circles encompass fractions $I_{\text{red}} = 0.3$, respectively.}
\label{fig:foc-transport}
\end{figure}

Ultimately, NMO lens systems following the prototype design are most likely not useful for high-resolution neutron imaging. However, they may still be useful for imaging at intermediate length scales, for neutron scattering investigations of small samples, and of samples exposed to extreme conditions, such as high pressure and high magnetic fields, and for adapting the beam size during neutron transport. Referring to the transport system illustrated in Fig. \ref{fig:Neutron-Beam-Transport} (a) ($f_1 = \SI{6}{\meter}$), a variation of the second NMO's focal length between $\SI{2}{\meter}\leq f_2 \leq \SI{6}{\meter}$, generates smaller beam sizes between $\SI{1}{\centi\meter}$ and $\SI{3}{\centi\meter}$ at F$_2$ from a circular moderator with diameter $\diameter_{\text{mod}} = \SI{3}{\centi\meter}$ at F$_1$ (compare Fig. \ref{fig:foc-transport}). Future technological developments may provide clues on the feasibility of realizing NMO with high performance for applications involving length scales on the order of micrometers.\\

\newpage
\section{Conclusions and outlook}
\label{sec:conclusionOutlook}

The reported experimental results, obtained with the small prototype elliptic nested mirror optics (NMO) equipped with polarizing $m = 4.1$ supermirrors, have demonstrated a high figure of merit for imaging, as defined by Eq.\,\ref{eq:FoM}, of \SI{72}{\percent}. The demonstration of unblurred focusing of sub-millimeter wide beams was limited only by the spatial resolution of the detector, \SI{2.5}{\milli\meter}. Future experiments will extend these studies to larger elliptic and parabolic NMO, and make use of higher-resolution detectors.\\

NMO are well-suited to the extraction of neutrons from a small moderator. In the standard technique, which uses neutron guides, illumination losses increase with the neutron wavelength and scale inversely to the moderator size. NMO on the other hand, can, by design, provide a high extraction efficiency, even for large wavelengths and especially for small beam sizes. As such, their natural field of application is in the development of delivery systems for the extraction of cold neutrons from compact, high-brilliance sources, like the flat para-hydrogen moderator at the ESS \cite{Andersen2018,Zanini2019} or tube-like moderators proposed for future accelerator-based neutron sources \cite{Habs2011,Gutberlet2019}. NMO are a viable alternative to non-linearly tapered neutron guides and Montel mirrors at such sources.\\

For neutron delivery to scattering instruments, NMO have the advantage that, in contrast to neutron guides, the phase space can be precisely selected and matched to the requirements of individual experiments. The imaging properties lead to a clean spatial definition of the beam, avoiding the penumbra that occurs behind neutron guides. The beam spectrum transmitted by an NMO is governed by geometry and the choice of the supermirror $m$-values. The well-defined angles of reflection in an NMO lead to a short-wavelength cut-off, whereas a neutron guide transports faster neutrons, so that additional devices for spectral cleaning are often needed. Flexibility in the choice of the beam divergence by remote-control of apertures can be used to optimize the signal-to-background ratio, e.g., when operating an instrument in high-resolution or high intensity mode, or when matching the beam to the acceptance of a sample environment. Since the apertures are located far from the experiment, associated backgrounds are small. This arrangement also leaves plenty of space for advanced sample environments.\\

NMO are also simpler to install than neutron guides. Their neutron extraction performance increases with NMO size and, hence, with distance to the source (up to distances exceeding several tens of meters, at which point neutron trajectories are appreciably curved by gravity). Increased distance from the source carries the additional benefit that irradiation damage to and activation of the optics are strongly reduced, the latter of which also means that NMO can be easily accessed and exchanged to accommodate varying needs of beam lines. From a technological point of view, the complexity of producing NMO is comparable to the manufacture of neutron benders \cite{schaerpfPropertiesBeamBender1989}, which are frequently used to place instruments out of the direct line-of-sight of a neutron source.\\

NMO can be used in a variety of ways. Their complementary neutron transport and imaging focusing capabilities not only enable the configuration of entire beam lines for dedicated purposes, but are especially suitable for satisfying instrument needs. Long beam lines, which transport neutrons over more than 100 m, as well as small optical setups, e.g., for focusing existing beams onto small samples, are possible.\\

Owing to their superior large-wavelength extraction capability, NMO will be an asset for future sources of very-cold neutrons (VCN), for which new moderator materials are being studied (see, e.g., Ref.\,\cite{Zimmer2016}). Higher neutron intensities at larger wavelengths would, for different classes of scattering instruments, lead to large gains of performance \cite{Micklich2005}. Experiments in fundamental-physics, including in-beam searches for a non-vanishing neutron electric dipole moment \cite{Piegsa2013} and searches for a baryon-number violation by neutron oscillations to antineutrons or sterile neutrons \cite{Addazi2021}, would also profit. NMO could be employed in such dedicated, large-scale projects, but also in general-purpose fundamental-physics beam lines, such as ANNI at the ESS \cite{Soldner2019}, or at in-beam sources of ultracold neutrons \cite{Pendlebury2014, Zimmer2011}.\\


\section*{Acknowledgement}
The authors are very grateful to Dr. Thomas Neulinger for careful scrutiny and many helpful comments on the manuscript. This work has been funded by the BMBF grant (Project No. 05K19WO3, MIEZEFOC) and by the Deutsche Forschungsgemeinschaft (DFG) within the Transregional Collaborative Research Center TRR 80 “From electronic correlations to functionality” - Projekt ID 107745057 - which is gratefully acknowledged. The presented experiments were performed at the MIRA instrument at Heinz Maier-Leibnitz Zentrum (MLZ), Garching, Germany.

\newpage
\bibliographystyle{unsrtnat_nourl}
\bibliography{Nested_Optic_Paper_lib}

\begin{thebibliography}{50}
\providecommand{\natexlab}[1]{#1}
\providecommand{\url}[1]{\texttt{#1}}
\expandafter\ifx\csname urlstyle\endcsname\relax
  \providecommand{\doi}[1]{doi: #1}\else
  \providecommand{\doi}{doi: \begingroup \urlstyle{rm}\Url}\fi

\bibitem[Shirane et~al.(2002)Shirane, Shapiro, and
  Tranquada]{shiraneNeutronScatteringTripleAxis2002}
Gen Shirane, Stephen~M. Shapiro, and John~M. Tranquada.
\newblock \emph{Neutron {Scattering} with a {Triple}-{Axis} {Spectrometer}:
  {Basic} {Techniques}}.
\newblock Cambridge University Press, 2002.
\newblock \doi{10.1017/CBO9780511534881}.

\bibitem[Guo et~al.(2012)Guo, Chen, Dai, Zhang, Guo, Chen, Wu, Gu, Gao, Yang,
  Yang, Dai, Mao, Sun, and Zhao]{guoPressureDrivenQuantumCriticality2012}
Jing Guo, Xiao-Jia Chen, Jianhui Dai, Chao Zhang, Jiangang Guo, Xiaolong Chen,
  Qi~Wu, Dachun Gu, Peiwen Gao, Lihong Yang, Ke~Yang, Xi~Dai, Ho-kwang Mao,
  Liling Sun, and Zhongxian Zhao.
\newblock Pressure-{Driven} {Quantum} {Criticality} in {Iron}-{Selenide}
  {Superconductors}.
\newblock \emph{Phys. Rev. Lett.}, 108\penalty0 (19):\penalty0 197001, May
  2012.
\newblock \doi{10.1103/PhysRevLett.108.197001}.
\newblock Publisher: American Physical Society.

\bibitem[Kobayashi et~al.(2002)Kobayashi, Hanazono, Tateiwa, Amaya, Haga,
  Settai, and Onuki]{kobayashiPressureinducedSuperconductivityFerromagnet2002}
T.~C. Kobayashi, K.~Hanazono, N.~Tateiwa, K.~Amaya, Y.~Haga, R.~Settai, and
  Y.~Onuki.
\newblock Pressure-induced superconductivity in a ferromagnet, {UGe2}:
  resistivity measurements in a magnetic field.
\newblock \emph{J. Phys.: Condens. Matter}, 14\penalty0 (44):\penalty0
  10779--10782, October 2002.
\newblock ISSN 0953-8984.
\newblock \doi{10.1088/0953-8984/14/44/376}.
\newblock Publisher: IOP Publishing.

\bibitem[Pfleiderer(2009)]{pfleidererSuperconductingPhasesElectron2009}
Christian Pfleiderer.
\newblock Superconducting phases of $f$-electron compounds.
\newblock \emph{Rev. Mod. Phys.}, 81\penalty0 (4):\penalty0 1551--1624,
  November 2009.
\newblock \doi{10.1103/RevModPhys.81.1551}.
\newblock Publisher: American Physical Society.

\bibitem[Fermi et~al.(1946)Fermi, Zinn, Laboratory, and
  Commission]{fermiReflectionNeutronsMirrors1946}
E.~Fermi, W.H. Zinn, Los Alamos~National Laboratory, and U.~S. Atomic~Energy
  Commission.
\newblock \emph{Reflection of {Neutrons} on {Mirrors}}.
\newblock United {States}. {Atomic} {Energy} {Commission}. {MDDC}. Manhattan
  District, 1946.

\bibitem[Maier-Leibnitz and
  Springer(1963)]{maier-leibnitzUseNeutronOptical1963}
H.~Maier-Leibnitz and T.~Springer.
\newblock The use of neutron optical devices on beam-hole experiments on
  beam-hole experiments.
\newblock \emph{Journal of Nuclear Energy. Parts A/B. Reactor Science and
  Technology}, 17\penalty0 (4):\penalty0 217 -- 225, 1963.
\newblock ISSN 0368-3230.
\newblock \doi{https://doi.org/10.1016/0368-3230(63)90022-3}.

\bibitem[Mezei(1976)]{mezeiNovelPolarizedNeutron1976a}
F.~Mezei.
\newblock Novel polarized neutron devices: supermirror and spin component
  amplifier.
\newblock \emph{Communications on Physics (London)}, 1\penalty0 (3):\penalty0
  81--85, 1976.

\bibitem[Mezei(1992)]{mezeiPolarizingSupermirrorDevices1992}
Ferenc Mezei.
\newblock Polarizing supermirror devices: some new developments.
\newblock In Charles~F. Majkrzak and James~L. Wood, editors, \emph{Neutron
  {Optical} {Devices} and {Applications}}, volume 1738, pages 107 -- 115. SPIE,
  1992.
\newblock \doi{10.1117/12.130623}.
\newblock Backup Publisher: International Society for Optics and Photonics.

\bibitem[Schanzer et~al.(2016)Schanzer, Schneider, and
  Böni]{schanzerNeutronOpticsApplications2016}
C.~Schanzer, M.~Schneider, and P.~Böni.
\newblock Neutron {Optics}: {Towards} {Applications} for {Hot} {Neutrons}.
\newblock \emph{Journal of Physics: Conference Series}, 746\penalty0
  (1):\penalty0 012024, 2016.

\bibitem[Aschauer et~al.(2000)Aschauer, Fleischmann, Schanzer, and
  Steichele]{aschauerNeutronGuidesFRMII2000}
H.~Aschauer, A.~Fleischmann, C.~Schanzer, and E.~Steichele.
\newblock Neutron guides at the {FRM}-{II}.
\newblock \emph{Physica B: Condensed Matter}, 283\penalty0 (4):\penalty0 323 --
  329, 2000.
\newblock ISSN 0921-4526.
\newblock \doi{https://doi.org/10.1016/S0921-4526(00)00324-0}.

\bibitem[Schanzer et~al.(2004)Schanzer, Böni, Filges, and
  Hils]{schanzerAdvancedGeometriesBallistic2004}
Christian Schanzer, Peter Böni, Uwe Filges, and Thomas Hils.
\newblock Advanced geometries for ballistic neutron guides.
\newblock \emph{Nuclear Instruments and Methods in Physics Research Section A:
  Accelerators, Spectrometers, Detectors and Associated Equipment},
  529\penalty0 (1):\penalty0 63 -- 68, 2004.
\newblock ISSN 0168-9002.
\newblock \doi{https://doi.org/10.1016/j.nima.2004.04.178}.

\bibitem[Mezei()]{Mezei1997}
F~Mezei.
\newblock The {Raisond}’{Etreof} {Long} {Pulse} {Spallation} {Sources}.
\newblock page~30.

\bibitem[Häse et~al.(2002)Häse, Knöpfler, Fiederer, Schmidt, Dubbers, and
  Kaiser]{Haese2002}
H.~Häse, A.~Knöpfler, K.~Fiederer, U.~Schmidt, D.~Dubbers, and W.~Kaiser.
\newblock A long ballistic supermirror guide for cold neutrons at {ILL}.
\newblock \emph{Nuclear Instruments and Methods in Physics Research Section A:
  Accelerators, Spectrometers, Detectors and Associated Equipment},
  485\penalty0 (3):\penalty0 453--457, June 2002.
\newblock ISSN 01689002.
\newblock \doi{10.1016/S0168-9002(01)02105-2}.

\bibitem[Zendler et~al.(2015)Zendler, Rodriguez, and
  Bentley]{zendlerGenericGuideConcepts2015}
C.~Zendler, D.~Martin Rodriguez, and P.~M. Bentley.
\newblock Generic guide concepts for the {European} {Spallation} {Source}.
\newblock \emph{Nuclear Instruments and Methods in Physics Research Section A:
  Accelerators, Spectrometers, Detectors and Associated Equipment},
  803:\penalty0 89 -- 99, 2015.
\newblock ISSN 0168-9002.
\newblock \doi{https://doi.org/10.1016/j.nima.2015.09.035}.

\bibitem[Rodriguez et~al.(2011)Rodriguez, Kennedy, and
  Bentley]{rodriguezPropertiesEllipticalGuides2011}
Damian~Martin Rodriguez, Shane~J. Kennedy, and Phillip~M. Bentley.
\newblock Properties of elliptical guides for neutron beam transport and
  applications for new instrumentation concepts.
\newblock \emph{Journal of Applied Crystallography}, 44\penalty0 (4):\penalty0
  727--737, August 2011.
\newblock \doi{10.1107/S0021889811018590}.

\bibitem[Stahn et~al.(2011)Stahn, Panzner, Filges, Marcelot, and
  Böni]{Stahn2011}
J.~Stahn, T.~Panzner, U.~Filges, C.~Marcelot, and P.~Böni.
\newblock Study on a focusing, low-background neutron delivery system.
\newblock \emph{Nuclear Instruments and Methods in Physics Research Section A:
  Accelerators, Spectrometers, Detectors and Associated Equipment},
  634\penalty0 (1, Supplement):\penalty0 S12--S16, 2011.
\newblock ISSN 0168-9002.
\newblock \doi{https://doi.org/10.1016/j.nima.2010.06.221}.

\bibitem[Cussen et~al.(2013)Cussen, Nekrassov, Zendler, and
  Lieutenant]{cussenMultipleReflectionsElliptic2013}
L.~D. Cussen, D.~Nekrassov, C.~Zendler, and K.~Lieutenant.
\newblock Multiple reflections in elliptic neutron guide tubes.
\newblock \emph{Nuclear Instruments and Methods in Physics Research Section A:
  Accelerators, Spectrometers, Detectors and Associated Equipment},
  705:\penalty0 121--131, 2013.
\newblock ISSN 0168-9002.
\newblock \doi{https://doi.org/10.1016/j.nima.2012.11.183}.

\bibitem[Weichselbaumer et~al.(2015)Weichselbaumer, Brandl, Georgii, Stahn,
  Panzner, and Böni]{Weichselbaumer2015}
S.~Weichselbaumer, G.~Brandl, R.~Georgii, J.~Stahn, T.~Panzner, and P.~Böni.
\newblock Tailoring phase-space in neutron beam extraction.
\newblock \emph{Nuclear Instruments and Methods in Physics Research Section A:
  Accelerators, Spectrometers, Detectors and Associated Equipment},
  793:\penalty0 75--80, 2015.
\newblock ISSN 0168-9002.
\newblock \doi{https://doi.org/10.1016/j.nima.2015.05.003}.

\bibitem[Zimmer(2018)]{zimmerImagingNestedmirrorAssemblies2018}
Oliver Zimmer.
\newblock Imaging nested-mirror assemblies – {A} new generation of neutron
  delivery systems?
\newblock \emph{Journal of Neutron Research}, 20:\penalty0 91--98, 2018.
\newblock ISSN 1477-2655.
\newblock \doi{10.3233/JNR-190101}.
\newblock Publisher: IOS Press.

\bibitem[Zimmer(2016{\natexlab{a}})]{zimmerMultimirrorImagingOptics2016}
Oliver Zimmer.
\newblock \emph{Multi-mirror imaging optics for low-loss transport of divergent
  neutron beams and tailored wavelength spectra}.
\newblock 2016{\natexlab{a}}.
\newblock \_eprint: 1611.07353.

\bibitem[Khaykovich et~al.(2011)Khaykovich, Gubarev, Bagdasarova, Ramsey, and
  Moncton]{khaykovichXrayTelescopesNeutron2011}
B.~Khaykovich, M.~V. Gubarev, Y.~Bagdasarova, B.~D. Ramsey, and D.~E. Moncton.
\newblock From x-ray telescopes to neutron scattering: {Using} axisymmetric
  mirrors to focus a neutron beam.
\newblock \emph{Nuclear Instruments and Methods in Physics Research Section A:
  Accelerators, Spectrometers, Detectors and Associated Equipment},
  631\penalty0 (1):\penalty0 98--104, March 2011.
\newblock ISSN 0168-9002.
\newblock \doi{10.1016/j.nima.2010.11.110}.

\bibitem[Wu et~al.(2019)Wu, Yang, Hussey, Wang, Song, Zhang, Wang, Wang, and
  Wang]{Wu2019}
Huarui Wu, Yang Yang, Daniel~S. Hussey, Zhiyuan Wang, Kun Song, Zhong Zhang,
  Zhanshan Wang, Zhe Wang, and Xuewu Wang.
\newblock Study of a nested neutron-focusing supermirror system for small-angle
  neutron scattering.
\newblock \emph{Nuclear Instruments and Methods in Physics Research Section A:
  Accelerators, Spectrometers, Detectors and Associated Equipment},
  940:\penalty0 380--386, 2019.
\newblock ISSN 0168-9002.
\newblock \doi{https://doi.org/10.1016/j.nima.2019.06.054}.

\bibitem[Georgii et~al.(2018)Georgii, Weber, Brandl, Skoulatos, Janoschek,
  Mühlbauer, Pfleiderer, and
  Böni]{georgiiMultipurposeThreeaxisSpectrometer2018}
R.~Georgii, T.~Weber, G.~Brandl, M.~Skoulatos, M.~Janoschek, S.~Mühlbauer,
  C.~Pfleiderer, and P.~Böni.
\newblock The multi-purpose three-axis spectrometer ({TAS}) {MIRA} at {FRM}
  {II}.
\newblock \emph{Nuclear Instruments and Methods in Physics Research Section A:
  Accelerators, Spectrometers, Detectors and Associated Equipment},
  881:\penalty0 60 -- 64, 2018.
\newblock ISSN 0168-9002.
\newblock \doi{https://doi.org/10.1016/j.nima.2017.09.063}.

\bibitem[Köhli et~al.(2016{\natexlab{a}})Köhli, Klein, Allmendinger,
  Perrevoort, Schröder, Martin, Schmidt, and
  Schmidt]{kohliCASCADEMultilayerBoron102016}
M.~Köhli, M.~Klein, F.~Allmendinger, A.-K. Perrevoort, T.~Schröder,
  N.~Martin, C.~J. Schmidt, and U.~Schmidt.
\newblock {CASCADE} - a multi-layer {Boron}-10 neutron detection system.
\newblock \emph{Journal of Physics: Conference Series}, 746:\penalty0 012003,
  September 2016{\natexlab{a}}.
\newblock \doi{10.1088/1742-6596/746/1/012003}.
\newblock Publisher: IOP Publishing.

\bibitem[Köhli et~al.(2016{\natexlab{b}})Köhli, Allmendinger, Häußler,
  Schröder, Klein, Meven, and Schmidt]{kohliEfficiencySpatialResolution2016}
M.~Köhli, F.~Allmendinger, W.~Häußler, T.~Schröder, M.~Klein, M.~Meven, and
  U.~Schmidt.
\newblock Efficiency and spatial resolution of the {CASCADE} thermal neutron
  detector.
\newblock \emph{Nuclear Instruments and Methods in Physics Research Section A:
  Accelerators, Spectrometers, Detectors and Associated Equipment},
  828:\penalty0 242--249, August 2016{\natexlab{b}}.
\newblock ISSN 01689002.
\newblock \doi{10.1016/j.nima.2016.05.014}.

\bibitem[Klenø et~al.(2012)Klenø, Lieutenant, Andersen, and
  Lefmann]{klenoSystematicPerformanceStudy2012}
Kaspar~Hewitt Klenø, Klaus Lieutenant, Ken~H. Andersen, and Kim Lefmann.
\newblock Systematic performance study of common neutron guide geometries.
\newblock \emph{Nuclear Instruments and Methods in Physics Research Section A:
  Accelerators, Spectrometers, Detectors and Associated Equipment},
  696:\penalty0 75--84, December 2012.
\newblock ISSN 01689002.
\newblock \doi{10.1016/j.nima.2012.08.027}.

\bibitem[Willendrup and Lefmann(2020)]{willendrupMcStasIntroductionUse2020}
Peter~Kjær Willendrup and Kim Lefmann.
\newblock {McStas} (i): {Introduction}, use, and basic principles for
  ray-tracing simulations.
\newblock \emph{Journal of Neutron Research}, 22\penalty0 (1):\penalty0 1--16,
  January 2020.
\newblock ISSN 1023-8166.
\newblock \doi{10.3233/JNR-190108}.
\newblock Publisher: IOS Press.

\bibitem[Andersen et~al.(2018)Andersen, Bertelsen, Zanini, Klinkby,
  Schönfeldt, Bentley, and Saroun]{Andersen2018}
Ken~Holst Andersen, Mads Bertelsen, Luca Zanini, Esben~Bryndt Klinkby, Troels
  Schönfeldt, Phillip~Martin Bentley, and Jan Saroun.
\newblock Optimization of moderators and beam extraction at the {ESS}.
\newblock \emph{Journal of Applied Crystallography}, 51\penalty0 (2):\penalty0
  264--281, April 2018.
\newblock \doi{10.1107/S1600576718002406}.

\bibitem[Zanini et~al.(2019)Zanini, Andersen, Batkov, Klinkby, Mezei,
  Schönfeldt, and Takibayev]{Zanini2019}
L.~Zanini, K.H. Andersen, K.~Batkov, E.B. Klinkby, F.~Mezei, T.~Schönfeldt,
  and A.~Takibayev.
\newblock Design of the cold and thermal neutron moderators for the {European}
  {Spallation} {Source}.
\newblock \emph{Nuclear Instruments and Methods in Physics Research Section A:
  Accelerators, Spectrometers, Detectors and Associated Equipment},
  925:\penalty0 33--52, 2019.
\newblock ISSN 0168-9002.
\newblock \doi{https://doi.org/10.1016/j.nima.2019.01.003}.

\bibitem[Cronert et~al.(2016)Cronert, Dabruck, Doege, Bessler, Klaus, Hofmann,
  Zakalek, Rücker, Lange, Butzek, Hansen, Nabbi, and Brückel]{cronert2016}
T~Cronert, J~P Dabruck, P~E Doege, Y~Bessler, M~Klaus, M~Hofmann, P~Zakalek,
  U~Rücker, C~Lange, M~Butzek, W~Hansen, R~Nabbi, and T~Brückel.
\newblock High brilliant thermal and cold moderator for the {HBS} neutron
  source project {Jülich}.
\newblock \emph{J. Phys.: Conf. Ser.}, 746:\penalty0 012036, September 2016.
\newblock ISSN 1742-6588, 1742-6596.
\newblock \doi{10.1088/1742-6596/746/1/012036}.

\bibitem[Carpenter(2019)]{carpenterDevelopmentCompactNeutron2019}
John~M. Carpenter.
\newblock The development of compact neutron sources.
\newblock \emph{Nature Reviews Physics}, 1\penalty0 (3):\penalty0 177--179,
  March 2019.
\newblock ISSN 2522-5820.
\newblock \doi{10.1038/s42254-019-0024-8}.
\newblock Number: 3 Publisher: Nature Publishing Group.

\bibitem[Schönfeldt et~al.(2013)Schönfeldt, Batkov, Klinkby, Lauritzen,
  Mezei, Pitcher, Takibayev, Willendrup, and Zanini]{schonfeldt2013}
Troels Schönfeldt, K.~Batkov, Esben~Bryndt Klinkby, Bent Lauritzen, F.~Mezei,
  E.~Pitcher, A.~Takibayev, Peter~Kjær Willendrup, and L.~Zanini.
\newblock Optimization of cold neutron beam extraction at {ESS}.
\newblock 2013.

\bibitem[Schneider et~al.(2009)Schneider, Stahn, and
  Böni]{schneiderFocusingColdNeutrons2009}
M.~Schneider, J.~Stahn, and P.~Böni.
\newblock Focusing of cold neutrons: {Performance} of a laterally graded and
  parabolically bent multilayer.
\newblock \emph{Nuclear Instruments and Methods in Physics Research Section A:
  Accelerators, Spectrometers, Detectors and Associated Equipment},
  610\penalty0 (2):\penalty0 530--533, November 2009.
\newblock ISSN 0168-9002.
\newblock \doi{10.1016/j.nima.2009.08.047}.

\bibitem[Schaerpf(1989)]{schaerpfPropertiesBeamBender1989}
O.~Schaerpf.
\newblock Properties of beam bender type neutron polarizers using supermirrors.
\newblock \emph{Physica B: Condensed Matter}, 156-157:\penalty0 639 -- 646,
  1989.
\newblock ISSN 0921-4526.
\newblock \doi{https://doi.org/10.1016/0921-4526(89)90751-5}.

\bibitem[Alefeld et~al.(1965)Alefeld, Christ, Schmatz, Kukla, and
  Scherm]{alefeldNeutronenleiterBerichtUber1965}
B.~Alefeld, J.~Christ, W.~Schmatz, D.~Kukla, and R.~Scherm.
\newblock Neutronenleiter : ein {Bericht} über den derzeitigen {Stand} der
  {Entwicklung}.
\newblock Technical Report FZJ-2017-03252, Kernforschungsanlage Jülich,
  Verlag, 1965.

\bibitem[Böni(2014)]{boeni2014}
Peter Böni.
\newblock High intensity neutron beams for small samples.
\newblock \emph{Journal of Physics: Conference Series}, 502:\penalty0 012047,
  April 2014.
\newblock \doi{10.1088/1742-6596/502/1/012047}.
\newblock Publisher: IOP Publishing.

\bibitem[Stahn and Glavic(2016)]{stahnFocusingNeutronReflectometry2016}
J.~Stahn and A.~Glavic.
\newblock Focusing neutron reflectometry: {Implementation} and experience on
  the {TOF}-reflectometer {Amor}.
\newblock \emph{Nuclear Instruments and Methods in Physics Research Section A:
  Accelerators, Spectrometers, Detectors and Associated Equipment},
  821:\penalty0 44--54, June 2016.
\newblock ISSN 0168-9002.
\newblock \doi{10.1016/j.nima.2016.03.007}.

\bibitem[Schanzer et~al.(2018)Schanzer, Schneider, Filges, and
  Böni]{Schanzer2018}
Christian Schanzer, Michael Schneider, Uwe Filges, and Peter Böni.
\newblock Variable focusing system for neutrons.
\newblock \emph{Journal of Physics: Conference Series}, 1021:\penalty0 012023,
  May 2018.
\newblock \doi{10.1088/1742-6596/1021/1/012023}.
\newblock Publisher: IOP Publishing.

\bibitem[Hils et~al.(2004)Hils, Boeni, and Stahn]{hils2004}
T~Hils, P~Boeni, and J~Stahn.
\newblock Focusing parabolic guide for very small samples.
\newblock \emph{Physica B: Condensed Matter}, 350\penalty0 (1):\penalty0
  166--168, July 2004.
\newblock ISSN 0921-4526.
\newblock \doi{10.1016/j.physb.2004.04.020}.

\bibitem[Jorba et~al.(2018)Jorba, Schulz, Hussey, Abir, Seifert, Tsurkan,
  Loidl, Pfleiderer, and
  Khaykovich]{jorbaHighresolutionNeutronDepolarization2018}
Pau Jorba, Michael Schulz, Daniel~S. Hussey, Muhammad Abir, Marc Seifert,
  Vladimir Tsurkan, Alois Loidl, Christian Pfleiderer, and Boris Khaykovich.
\newblock High-resolution neutron depolarization microscopy of the
  ferromagnetic transitions in {Ni}$_3${Al} and {HgCr}$_2${Se}$_4$ under
  pressure.
\newblock \emph{arXiv:1812.00864 [cond-mat]}, December 2018.
\newblock \doi{10.1016/j.jmmm.2018.11.086}.
\newblock arXiv: 1812.00864.

\bibitem[Trtik et~al.(2015)Trtik, Hovind, Grünzweig, Bollhalder, Thominet,
  David, Kaestner, and Lehmann]{Trtik2015}
Pavel Trtik, Jan Hovind, Christian Grünzweig, Alex Bollhalder, Vincent
  Thominet, Christian David, Anders Kaestner, and Eberhard~H. Lehmann.
\newblock Improving the {Spatial} {Resolution} of {Neutron} {Imaging} at {Paul}
  {Scherrer} {Institut} – {The} {Neutron} {Microscope} {Project}.
\newblock \emph{Physics Procedia}, 69:\penalty0 169--176, January 2015.
\newblock ISSN 1875-3892.
\newblock \doi{10.1016/j.phpro.2015.07.024}.

\bibitem[Habs et~al.(2011)Habs, Gross, Thirolf, and Böni]{Habs2011}
D.~Habs, M.~Gross, P.~G. Thirolf, and P.~Böni.
\newblock Neutron halo isomers in stable nuclei and their possible application
  for the production of low energy, pulsed, polarized neutron beams of high
  intensity and high brilliance.
\newblock \emph{Applied Physics B}, 103:\penalty0 485--499, 2011.
\newblock \doi{https://doi.org/10.1007/s00340-010-4276-3}.

\bibitem[Gutberlet et~al.(2019)Gutberlet, Rücker, Zakalek, Cronert, Voigt,
  Baggemann, Doege, Mauerhofer, Böhm, Dabruck, Nabbi, Butzek, Klaus, Lange,
  and Brückel]{Gutberlet2019}
T.~Gutberlet, U.~Rücker, P.~Zakalek, T.~Cronert, J.~Voigt, J.~Baggemann, P.-E.
  Doege, E.~Mauerhofer, S.~Böhm, J.P. Dabruck, R.~Nabbi, M.~Butzek, M.~Klaus,
  C.~Lange, and T.~Brückel.
\newblock The {Jülich} high brilliance neutron source project – {Improving}
  access to neutrons.
\newblock \emph{Physica B: Condensed Matter}, 570:\penalty0 345--348, 2019.
\newblock ISSN 0921-4526.
\newblock \doi{https://doi.org/10.1016/j.physb.2018.01.019}.

\bibitem[Zimmer(2016{\natexlab{b}})]{Zimmer2016}
Oliver Zimmer.
\newblock Neutron conversion and cascaded cooling in paramagnetic systems for a
  high-flux source of very cold neutrons.
\newblock \emph{Phys. Rev. C}, 93\penalty0 (3):\penalty0 035503, March
  2016{\natexlab{b}}.
\newblock \doi{10.1103/PhysRevC.93.035503}.
\newblock Publisher: American Physical Society.

\bibitem[Micklich and Carpenter(2005)]{Micklich2005}
B.~J. Micklich and John~M. Carpenter.
\newblock Proceedings of the {Workshop} on {Applications} of a {Very} {Cold}
  {Neutron} {Source}.
\newblock In \emph{{ANL}-05/42}, 2005.

\bibitem[Piegsa(2013)]{Piegsa2013}
Florian~M. Piegsa.
\newblock New concept for a neutron electric dipole moment search using a
  pulsed beam.
\newblock \emph{Phys. Rev. C}, 88\penalty0 (4):\penalty0 045502, October 2013.
\newblock \doi{10.1103/PhysRevC.88.045502}.
\newblock Publisher: American Physical Society.

\bibitem[Addazi et~al.(2021)Addazi, Anderson, Ansell, Babu, Barrow, Baxter,
  Bentley, Berezhiani, Bevilacqua, Biondi, Bohm, Brooijmans, Broussard,
  Cedercäll, Crawford, Dev, DiJulio, Dolgov, Dunne, Fierlinger, Fitzsimmons,
  Fomin, Frost, Gardiner, Gardner, Galindo-Uribarri, Geltenbort, Girmohanta,
  Golubev, Golubeva, Greene, Greenshaw, Gudkov, Hall-Wilton, Heilbronn,
  Herrero-Garcia, Holley, Ichikawa, Ito, Iverson, Johansson, Jönsson, Jwa,
  Kamyshkov, Kanaki, Kearns, Kokai, Kerbikov, Kitaguchi, Kittelmann, Klinkby,
  Kobakhidze, Koerner, Kopeliovich, Kozela, Kudryavtsev, Kupsc, Lee, Lindroos,
  Makkinje, Marquez, Meirose, Miller, Milstead, Mohapatra, Morishima, Muhrer,
  Mumm, Nagamoto, Nepomuceno, Nesti, Nesvizhevsky, Nilsson, Oskarsson, Paryev,
  Pattie, Penttil, Perrey, Pokotilovski, Potashnikovav, Ramic, Redding,
  Richard, Ries, Rinaldi, Rizzi, Rossi, Ruggles, Rybolt, Santoro, Sarkar,
  Saunders, Senjanovic, Serebrov, Shimizu, Shrock, Silverstein, Silvermyr,
  Snow, Takibayev, Tkachev, Townsend, Tureanu, Varriano, Vainshtein, Vries,
  Wagner, Woracek, Yamagata, Yiu, Young, Zanini, Zhang, and Zimmer]{Addazi2021}
A.~Addazi, K.~Anderson, S.~Ansell, K.~S. Babu, J.~L. Barrow, D.~V. Baxter,
  P.~M. Bentley, Z.~Berezhiani, R.~Bevilacqua, R.~Biondi, C.~Bohm,
  G.~Brooijmans, L.~J. Broussard, J.~Cedercäll, C.~Crawford, P.~S.~B. Dev,
  D.~D. DiJulio, A.~D. Dolgov, K.~Dunne, P.~Fierlinger, M.~R. Fitzsimmons,
  A.~Fomin, M.~J. Frost, S.~Gardiner, S.~Gardner, A.~Galindo-Uribarri,
  P.~Geltenbort, S.~Girmohanta, P.~Golubev, E.~Golubeva, G.~L. Greene,
  T.~Greenshaw, V.~Gudkov, R.~Hall-Wilton, L.~Heilbronn, J.~Herrero-Garcia,
  A.~Holley, G.~Ichikawa, T.~M. Ito, E.~Iverson, T.~Johansson, L.~Jönsson,
  Y.-J. Jwa, Y.~Kamyshkov, K.~Kanaki, E.~Kearns, Z.~Kokai, B.~Kerbikov,
  M.~Kitaguchi, T.~Kittelmann, E.~Klinkby, A.~Kobakhidze, L.~W. Koerner,
  B.~Kopeliovich, A.~Kozela, V.~Kudryavtsev, A.~Kupsc, Y.~T. Lee, M.~Lindroos,
  J.~Makkinje, J.~I. Marquez, B.~Meirose, T.~M. Miller, D.~Milstead, R.~N.
  Mohapatra, T.~Morishima, G.~Muhrer, H.~P. Mumm, K.~Nagamoto, A.~Nepomuceno,
  F.~Nesti, V.~V. Nesvizhevsky, T.~Nilsson, A.~Oskarsson, E.~Paryev, R.~W.
  Pattie, S.~Penttil, H.~Perrey, Y.~N. Pokotilovski, I.~Potashnikovav,
  K.~Ramic, C.~Redding, J.-M. Richard, D.~Ries, E.~Rinaldi, N.~Rizzi, N.~Rossi,
  A.~Ruggles, B.~Rybolt, V.~Santoro, U.~Sarkar, A.~Saunders, G.~Senjanovic,
  A.~P. Serebrov, H.~M. Shimizu, R.~Shrock, S.~Silverstein, D.~Silvermyr, W.~M.
  Snow, A.~Takibayev, I.~Tkachev, L.~Townsend, A.~Tureanu, L.~Varriano,
  A.~Vainshtein, J.~de Vries, R.~Wagner, R.~Woracek, Y.~Yamagata, S.~Yiu, A.~R.
  Young, L.~Zanini, Z.~Zhang, and O.~Zimmer.
\newblock New high-sensitivity searches for neutrons converting into
  antineutrons and/or sterile neutrons at the {HIBEAM}/{NNBAR} experiment at
  the {European} {Spallation} {Source}.
\newblock \emph{J. Phys. G: Nucl. Part. Phys.}, 48\penalty0 (7):\penalty0
  070501, June 2021.
\newblock ISSN 0954-3899.
\newblock \doi{10.1088/1361-6471/abf429}.
\newblock Publisher: IOP Publishing.

\bibitem[Soldner et~al.(2019)Soldner, Abele, Konrad, Märkisch, Piegsa,
  Schmidt, Theroine, and Sánchez]{Soldner2019}
Torsten Soldner, Hartmut Abele, Gertrud Konrad, Bastian Märkisch, Florian~M.
  Piegsa, Ulrich Schmidt, Camille Theroine, and Pablo~Torres Sánchez.
\newblock {ANNI} – {A} pulsed cold neutron beam facility for particle physics
  at the {ESS}.
\newblock \emph{EPJ Web Conf.}, 219:\penalty0 10003, 2019.
\newblock ISSN 2100-014X.
\newblock \doi{10.1051/epjconf/201921910003}.
\newblock Publisher: EDP Sciences.

\bibitem[Pendlebury and Greene(2014)]{Pendlebury2014}
J.~M. Pendlebury and G.~L. Greene.
\newblock Considerations for an {Intense} {Source} of {Ultracold} {Neutrons} at
  the {European} {Long} {Pulse} {Spallation} {Source}.
\newblock \emph{Physics Procedia}, 51:\penalty0 78--84, January 2014.
\newblock ISSN 1875-3892.
\newblock \doi{10.1016/j.phpro.2013.12.018}.

\bibitem[Zimmer et~al.(2011)Zimmer, Piegsa, and Ivanov]{Zimmer2011}
Oliver Zimmer, Florian~M. Piegsa, and Sergey~N. Ivanov.
\newblock Superthermal {Source} of {Ultracold} {Neutrons} for {Fundamental}
  {Physics} {Experiments}.
\newblock \emph{Phys. Rev. Lett.}, 107\penalty0 (13):\penalty0 134801,
  September 2011.
\newblock \doi{10.1103/PhysRevLett.107.134801}.
\newblock Publisher: American Physical Society.

\end{thebibliography}
\newpage


\appendix


\section{Simulated NMO efficiencies for various mirror parameters}\label{app:simulation_optic}

In this appendix we discuss simulations carried out in order to study the influence of the substrate thickness $d_{\text{sub}}$ and the non-ideal mirror reflectivity on the neutron transport efficiencies, $Q$ (as defined in Section \ref{sec:meas_res}), in which the geometry of the NMO prototype and a monochromatic beam spectrum were used. Although, on account of their availability, only double-side coated mirrors were used in the NMO prototype (see Section \ref{sect:experimentalSetup}), simulations were performed with both single-side and double-side coated mirrors for comparison.\\ 

\begin{figure}[htb]
	\centering
	\includegraphics[width = \linewidth]{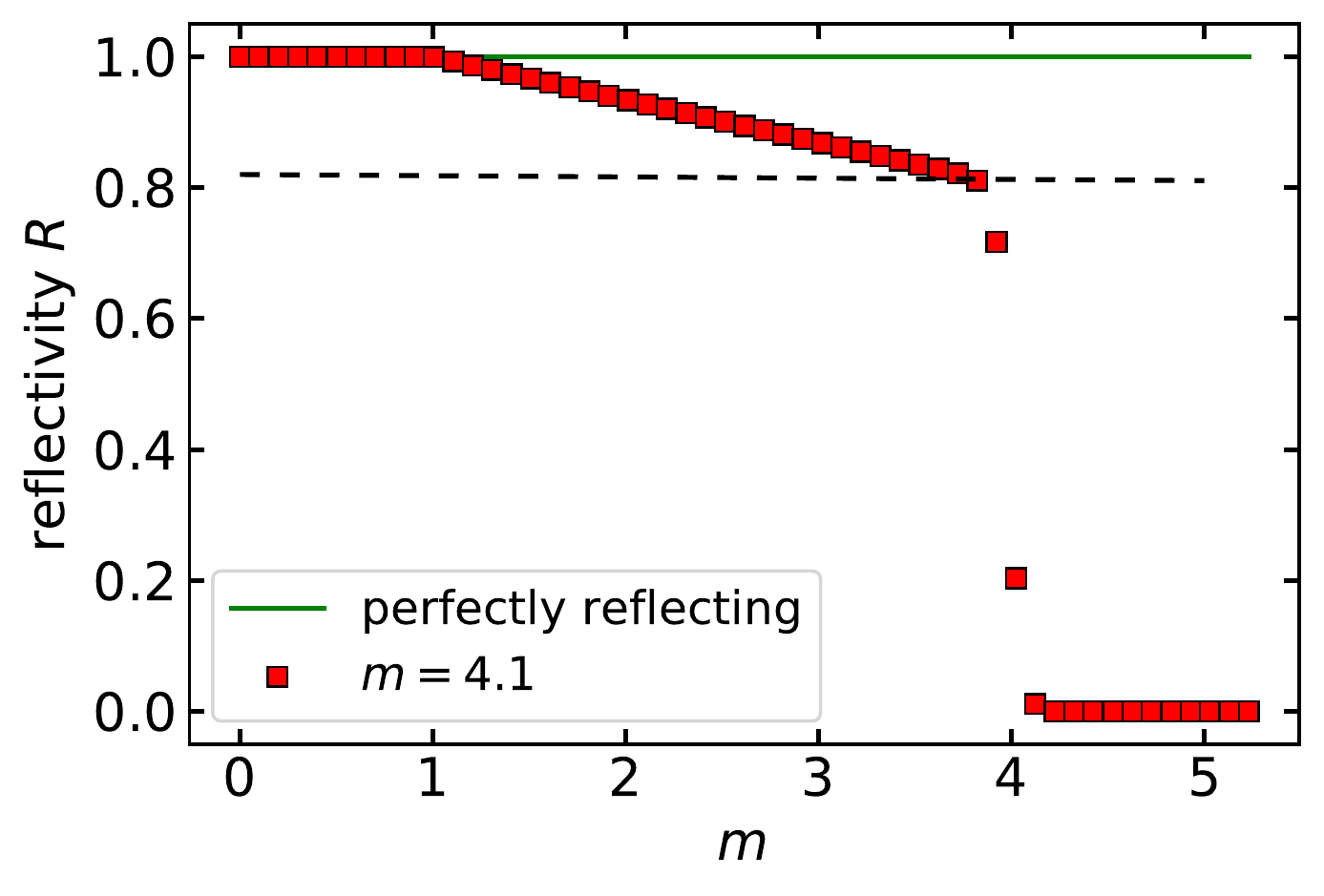}
	\caption{Reflectivity curves, approximating a measured $m = 4.1$ polarizing supermirror (red squares) and an ideal mirror (green line) with $R = 1.0$ for $0 \le m \le \infty$.}
	\label{fig:ref_profile}
\end{figure}

Table \ref{tab:efficiencies_app} summarizes the influence of $d_{\text{sub}}$ and $w$ on the transport efficiency. Two sets of calculations were performed using the data shown in Fig.\,\ref{fig:ref_profile}: $Q_{\text{ideal}}$ (value to the left vertical line in Table \ref{tab:efficiencies_app}), which used an ideal reflectivity curve (green line), and $Q_{\text{real}}$ (value to the right), which used the measured reflectivity curve of an $m=4.1$ supermirror (red squares). From the data in Table \ref{tab:efficiencies_app} it may be seen that the relative drop $(Q_{\text{real}} - Q_{\text{ideal}})/Q_{\text{ideal}}\approx \SI{5}{\percent}$ stays well below the $\SI{20}{\percent}$ reduction of the edge reflectivity of the $m=4.1$ supermirrors. This is to be expected for the small angles of reflection and the selected monochromatic beam of MIRA with $\lambda=\SI{4.9}{\angstrom}$. The gradual decrease in $Q$ with increasing $d_{\text{sub}}$ is a result of channeling and double reflection (see Fig.\,\ref{fig:channeling}), whereas the decrease in $Q$ with increasing $w$ is due to increased geometric losses in transporting neutrons from an extended source.\\

\begin{table}[htb]
	\centering
	\begin{adjustbox}{width=\columnwidth}
		\begin{tabular}{c|c c c}
			$w$ (mm)& $d_{\text{sub}} = 0$ mm & $d_{\text{sub}} = \SI{0.15}{\milli\meter}$& $d_{\text{sub}} = \SI{0.30}{\milli\meter}$  \\ \hline
			0.25 &      0.99\textbar0.93  &   0.85\textbar\textbf{0.82}  & 0.70\textbar 0.69\\ 
			0.5 &      0.99\textbar0.92  &   0.85\textbar \textbf{0.81}  & 0.70\textbar 0.69\\
			1    &      0.97\textbar0.91   &  0.85\textbar \textbf{0.81}  & 0.70\textbar 0.69\\ 
			2    &      0.94\textbar0.88   &  0.84\textbar \textbf{0.80}  & 0.69\textbar 0.67\\ 
			4	 &		0.89\textbar0.83   &  0.81\textbar \textbf{0.77}  & 0.69\textbar 0.67\\
			6    &      0.84\textbar0.79   &  0.77\textbar \textbf{0.73} & 0.67\textbar 0.66
	\end{tabular}
	\end{adjustbox}
	\caption{NMO efficiencies $Q(w=\SI{6}{\milli\meter},\, w_{\text{int}}=\SI{9}{\milli\meter})$, simulated for various values of Si-substrate thickness, $d_{\text{sub}}$, and beam width $w$. The first and the second entry in each cell ($Q_{\text{ideal}}|Q_{\text{real}}$) shows results obtained for perfectly reflecting mirrors and for a realistic $m=4.1$ reflectivity profile, respectively. Simulations were performed keeping the experimental geometry, including the aperture A$_2$. Except for vanishing substrate thickness, double-side coated mirrors were assumed. The bold column corresponds to the entries in Table\,\ref{tab:efficiencies_main}.}
	\label{tab:efficiencies_app}
\end{table}

Next, we compare the efficiency of NMO whose mirrors consist of silicon wafers coated, either on one or both sides, with an $m=4.1$ supermirror. Double-side coatings are frequently employed in multi-mirror bender devices. In contrast, one might expect that, because neutron transport in NMO relies on single reflections, single-side coatings would be a more appropriate choice than double-side coatings. However, for imperfect reflectivity, double-side coated Si-wafers may still increase NMO efficiency if the mirrors are sufficiently thin. Consider the following argument; neutrons that leak through the first coating into the substrate are very likely to be reflected by the second coating at the back side. In order to contribute to ``single reflection'' transport, such neutrons must be subsequently transmitted through the first coating. Despite the low probability of such an event, one might, nevertheless, expect a net gain in neutron transport.\\

With increasing $d_{\text{sub}}$, the front faces of the wafers are exposed to the beam entering the NMO with an increasing filling fraction, $\zeta$, such that this gain is quickly exhausted by the channeling and double reflection effects described in Fig.\,\ref{fig:channeling}. For our prototype, where five $\SI{0.15}{\milli\meter}$ thick mirror plates were illuminated through an $\SI{8}{\milli\meter}$ wide aperture, this fraction amounted to $\zeta=\SI{9.4}{\percent}$.\\

\begin{figure}[]
	\centering
	\includegraphics[width = \columnwidth]{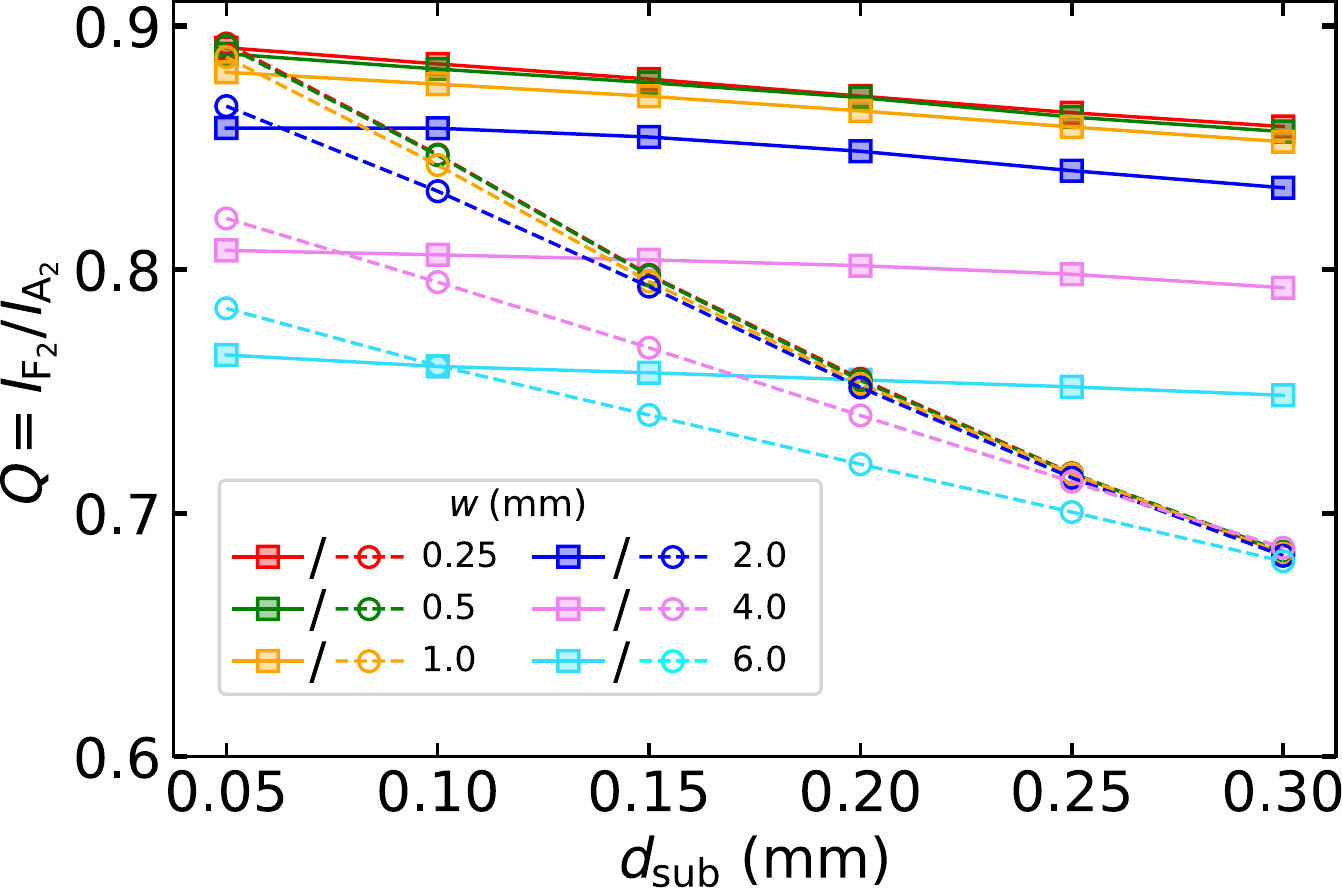}
	\caption{Simulated transport efficiencies, $Q$, for various combinations of $d_{\text{sub}}$ and $w$, for single-sided (filled squares, solid lines) and double-sided (empty circles, broken lines) $m=4.1$ supermirror coatings.}
	\label{fig:both_coated}
\end{figure}

Figure \ref{fig:both_coated} presents the results of McStas simulations for the NMO equipped with mirrors on both sides of the optical axis. Lacking the aperture A$_2$, but otherwise keeping the geometry of the experiment, the simulated data also include the divergence hole. They qualitatively show the dependences expected from the previous discussion. Notably, across all $w$, one observes only a weak dependence on $d_{\text{sub}}$ for mirrors with single-side coatings. The substrate thickness, below which a double-side coating provides a (for the chosen mirror reflectivity marginal) gain, is found to increase with the beam width $w$. The question of whether or not double-side coatings offer a sufficient advantage over single-side coatings has to be addressed in each concrete NMO design. It may be appropriate to use them in large NMO, where the filling fraction $\zeta$ can be kept small, or in the outer plates of NMO intended to cover a large solid angle, which requires supermirrors with atypically-large $m$ and, correspondingly, reduced edge reflectivity.\\

\section{Long-distance neutron transport from a circular moderator}
\label{app:pancake_extraction}

\begin{figure}[htb]
	\centering
	\includegraphics[width = \linewidth]{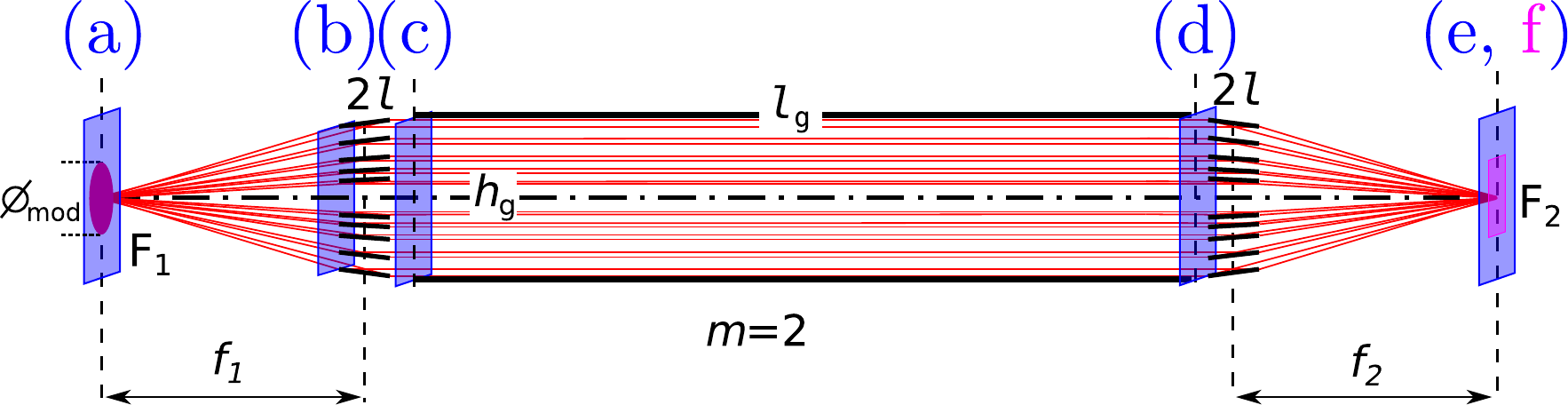}
	\caption{Sketch of the simulated long-distance neutron transport system based on two double-planar parabolic NMO connected by a long straight neutron guide. The violet quadrilaterals indicate PSDs used as beam monitors in the simulations. They are located at positions (a)-(e) and cover an area of $218\times\SI{218}{\milli\meter^2}$. The additional PSD (f) with an area of $30\times\SI{30}{\milli\meter^2}$ highlights the focused beam.}
	\label{fig:pancake_geometry}
\end{figure}

Here we present McStas simulation results of a combination of two double-planar parabolic NMO connected by a straight neutron guide, using example parameters that are not optimized for any specific purpose. (The operational principle of such a system was described in Section \ref{sec:transport neutrons}.) The simulation geometry is sketched in Fig.\,\ref{fig:pancake_geometry}. A circular moderator of diameter $\diameter_{\text{mod}} = \SI{30}{\milli\meter}$ illuminates the first NMO with an angle- and wavelength-independent brilliance, ranging between $\SI{2}{\angstrom}$ and $\SI{8}{\angstrom}$. Each double-planar NMO contains two subsystems, which act in the horizontal and vertical directions, respectively. The low-divergence beam produced by the first NMO enters a straight, $m=2$ neutron guide of length $l_{\text{g}} = \SI{160}{\meter}$ and square cross section with $\SI{218}{\milli\meter}$ on edge. The second NMO system refocuses the beam. The focal length is common to both NMO: $f_1 = f_2 = \SI{6}{\meter}$. The NMO are equipped with $m = 4.1$ supermirrors, have a total length of $2l=\SI{1.2}{\meter}$, and are designed to match to the cross section of the guide. The substrate thickness is neglected in the simulations, i.e., $d_{\text{sub}} = 0$.\\


\begin{figure}[htb]
		\centering
		\includegraphics[width = \linewidth]{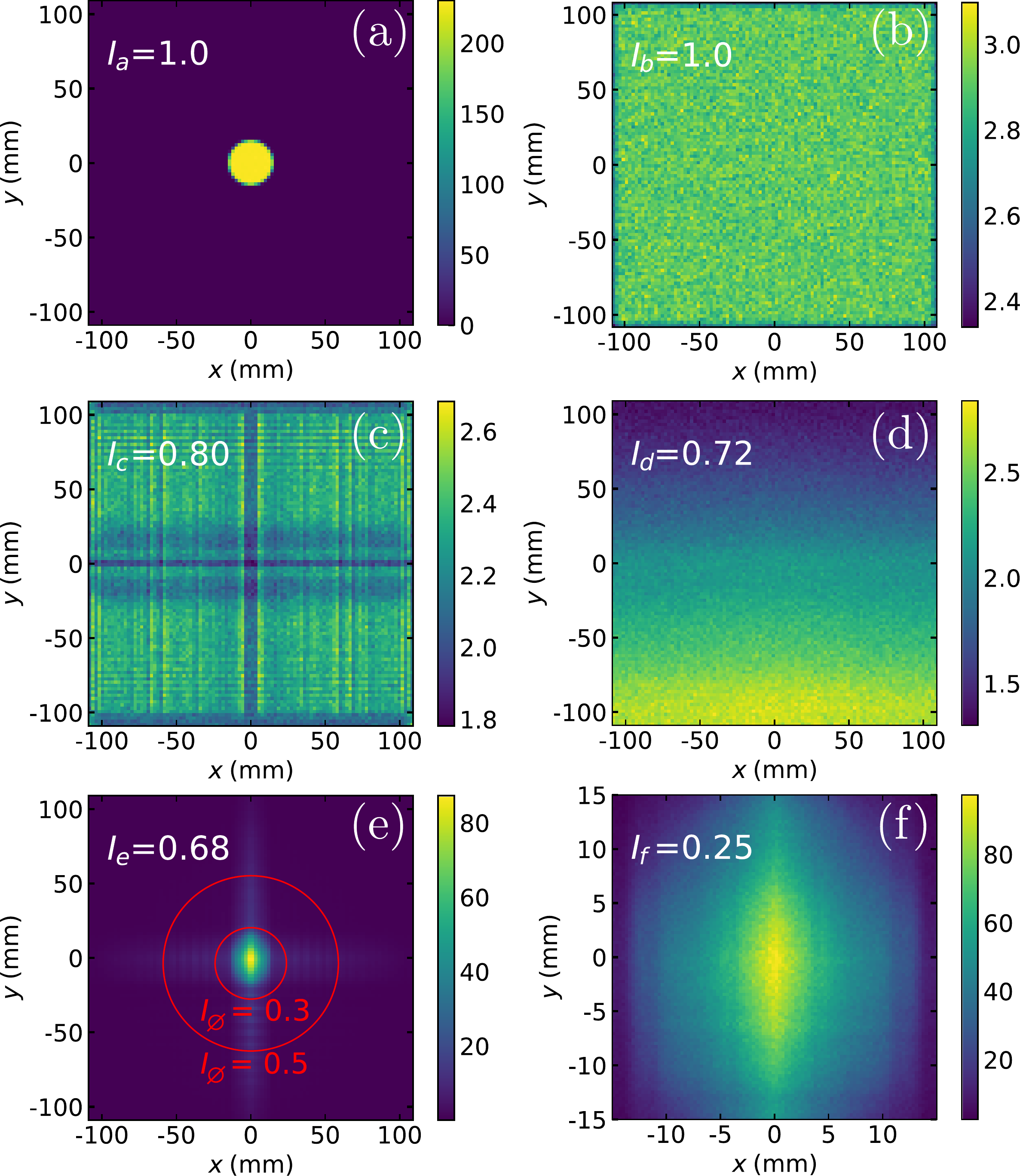}
		\caption{Simulated intensity distributions at the locations as labeled in Fig.\,\ref{fig:pancake_geometry}, including gravity. (a) Neutrons emitted from the circular moderator. (b) The homogeneously illuminated entrance of the first double-planar parabolic NMO. (c) At the exit of the NMO, a texture results from neutrons reflected at individual mirrors. (d) At the end of the guide, neutrons are found to be accumulated at the bottom, which is due to gravity. (e) and (f) The second NMO refocuses the neutrons. Red circles in (e) indicate areas encompassing integrated intensities of $I_{\diameter} = 0.3I_a$ and $I_{\diameter} = 0.5I_a$, respectively. About $\SI{25}{\percent}$ of the neutrons leaving the moderator in the direction of the first NMO are recovered within the area (f) of $30\times\SI{30}{\milli\meter^2}$. The fraction $I_j$ of neutrons arriving at the monitor (j) is shown in each corresponding plot. Intensities are normalized to the total intensity leaving the moderator in the direction of the NMO, i.e., $I_a = I_b = 1$.}
		\label{fig:pancake_overview}
\end{figure}

\begin{figure}[htb]
	\centering
	\includegraphics[width = \linewidth]{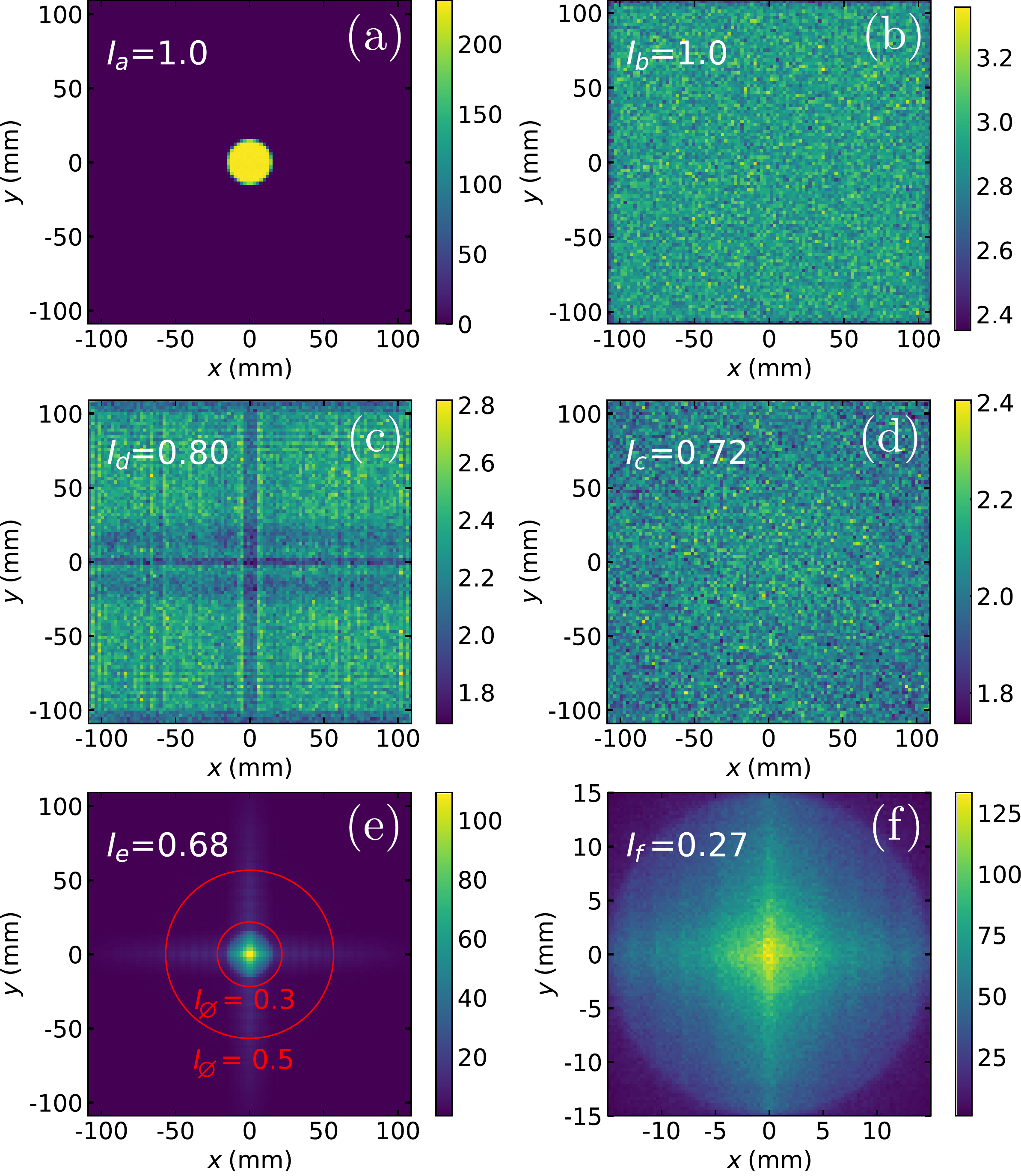}
	\caption{Intensity distributions obtained from simulations excluding gravity. A comparison with Fig.\,\ref{fig:pancake_overview} shows notably the absence of neutron accumulation in the lower part of the long guide (d), as well as of the vertical asymmetry in (f), both effects being attributed to gravity. Only $\SI{10}{\percent}$ of the neutrons are lost within the long guide, and the simulated intensities with and without gravity are in good agreement. This confirms the validity of this NMO-based concept for neutron extraction and transport over large distances.}
	\label{fig:pancake_overview_nog}
\end{figure}

Figures \ref{fig:pancake_overview} and \ref{fig:pancake_overview_nog} show simulated intensity distributions at the positions of the beam monitors (see Fig.\,\ref{fig:pancake_geometry}), with and without gravity, respectively. The texture in the intensity distribution at the beam monitor (c), after the first set of NMO, is due to the reflection of neutrons from the individual mirrors of the NMO, and the central, perpendicular stripes of lower intensity are a result of the divergence hole. Since the first planar subsystem in the beam has vertical mirrors, the horizontal stripe is more blurred than the vertical one. The gravity-induced large vertical gradient of neutron intensity at the end of the guide (d) is particularly pronounced for a low-divergence beam, as shown in complementary simulations. The texture visible in (c) averages away along the guide. $I_f=\SI{25}{\percent}$ of the neutrons illuminating the first NMO arrive within an area of $30\times\SI{30}{\milli\meter^2}$ at F$_2$. Gravity spreads the intensity distribution of the polychromatic beam with its maximum found marginally below $y=\SI{0}{\milli\meter}$.\\

\begin{figure}[htb]
	\centering
	\includegraphics[width = \linewidth]{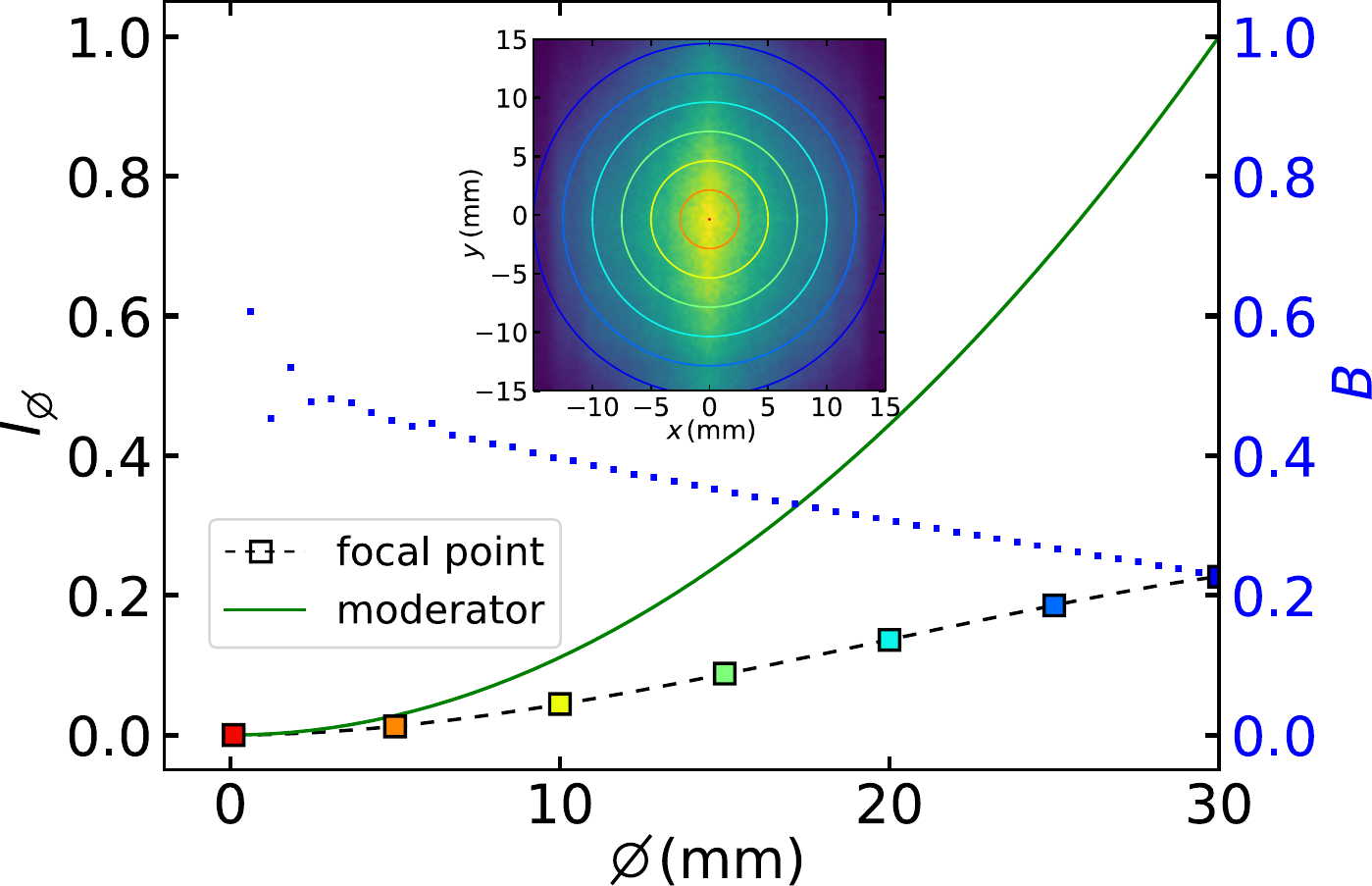}
	\caption{Intensity fraction and integrated brilliance transfer $B$, as measured with a circular monitor at F$_2$ with a diameter $\diameter$. The intensity data shown in the inset is the same as in Fig.\,\ref{fig:pancake_overview} (f). For comparison, the corresponding, quadratically rising intensity fractions at the moderator (a), with diameter $\SI{30}{\milli\meter}$, are shown in green. Values of $B$ shown in blue are larger than $\SI{40}{\percent}$ for a target area of $\diameter\leq \SI{10}{\milli\meter}$ and still $\SI{23}{\percent}$ for $\diameter=\diameter_{\text{mod}}=\SI{30}{\milli\meter}$. The colors of the circles match the colors of the data points.}
	\label{fig:int_vs_rad}
\end{figure}

Figure \ref{fig:int_vs_rad} shows the fraction of those neutrons entering the first NMO that finally arrive within a circle around the focal point of the second NMO, plotted against the circle diameter $\diameter$. For instance, circles of diameter $\diameter=\SI{10}{\milli\meter}$ and $\diameter=\SI{30}{\milli\meter}$ encompass intensity fractions $I_{\diameter} = 0.045$ and $I_{\diameter} = 0.23$, respectively. Note, however, that the integrated brilliance transfer, obtained according to its definition as the ratio of the intensity fraction to the moderator intensity, $B = I_{\diameter}/(\diameter/\diameter_{\text{mod}})^2$, increases with decreasing $\diameter$ and reaches $\approx \SI{50}{\percent}$ for a beam diameter of several millimeters. Noting also that beam losses of approximately 30\% due to finite supermirror reflectivity are already included in the simulations, these findings show an excellent performance of neutron transport to small samples. The results also show that the influence of gravity on the performance of such a system, comprised of compact NMO, is marginal.\\

\null
\vfill
\end{document}